\documentclass[10pt,letterpaper]{article}
\usepackage[top=0.85in,left=2.75in,footskip=0.75in]{geometry}

\usepackage[table]{xcolor}
\usepackage{array}
\usepackage{floatrow}
\usepackage{adjustbox}

\usepackage[T1]{fontenc}
\usepackage{lmodern}
\usepackage{microtype}
\usepackage{float}
\usepackage{caption}
\usepackage{booktabs}
\usepackage{blindtext}
\usepackage{amssymb}

\usepackage{amsmath,amssymb}

\usepackage{changepage}

\usepackage[utf8x]{inputenc}

\usepackage{textcomp,marvosym}

\usepackage{cite}

\usepackage{nameref,hyperref}

\usepackage[right]{lineno}

\usepackage{microtype}
\DisableLigatures[f]{encoding = *, family = * }

\usepackage[table]{xcolor}

\usepackage{array}

\newcolumntype{+}{!{\vrule width 2pt}}

\newlength\savedwidth

\raggedright
\usepackage[aboveskip=1pt,labelfont=bf,labelsep=period,justification=raggedright,singlelinecheck=off]{caption}

\makeatletter
\renewcommand{\@biblabel}[1]{\quad#1.}
\makeatother

\usepackage{lastpage,fancyhdr,graphicx}
\usepackage{epstopdf}
\fancyheadoffset[L]{2.25in}
\fancyfootoffset[L]{2.25in}
\begin{document}
\vspace*{0.2in}

\begin{flushleft}
{\Large
\textbf\newline{Artificial Intelligence-Based Analytics for Impacts of COVID-19 and Online Learning on College Students’ Mental Health} % Please use "sentence case" for title and headings (capitalize only the first word in a title (or heading), the first word in a subtitle (or subheading), and any proper nouns).
}
\newline
\\
Mostafa Rezapour\textsuperscript{1,*,\dag},
Scott K. Elmshaeuser\textsuperscript{1}
\\
\bigskip
\textbf{1} Department of Mathematics, Wake Forest University, Winston-Salem, NC, U.S.
\\
* rezapom@wfu.edu
\bigskip

\dag The corresponding author current address: Biomedical Informatics, Wake Forest University School of Medicine, Winston-Salem, NC, U.S. (mrezapou@wakehealth.edu)

% Group/Consortium Author Note

% Use the asterisk to denote corresponding authorship and provide email address in note below.

\end{flushleft}
\section*{Abstract}
COVID-19, the disease caused by the novel coronavirus (SARS-CoV-2), first emerged in Wuhan, China late in December 2019. Not long after, the virus spread worldwide and was declared a pandemic by the World Health Organization in March 2020. This caused many changes around the world and in the United States, including an educational shift towards online learning. In this paper, we seek to understand how the COVID-19 pandemic and increase in online learning impact college students’ emotional wellbeing. We use several machine learning and statistical models to analyze data collected by the Faculty of Public Administration at the University of Ljubljana, Slovenia in conjunction with an international consortium of universities, other higher education institutions, and students’ associations. Our results indicate that features related to students’ academic life have the largest impact on their emotional wellbeing. Other important factors include students’ satisfaction with their university’s and government’s handling of the pandemic as well as students’ financial security.

%\linenumbers

% Use "Eq" instead of "Equation" for equation citations.

\section*{Introduction}

COVID-19, the disease caused by the novel coronavirus (SARS-CoV-2), first emerged in Wuhan, China late in December 2019. Not long after, the virus began to spread worldwide, impacting the lives of millions. By January 30, 2020, the World Health Organization (WHO) declared the novel coronavirus to be an international health emergency, and by March 11, 2020, it was labeled a pandemic \cite{1}. In the United States, government and health care officials began recommending social distancing as a way to mitigate the spread of the disease, but eventually deemed it necessary to close schools and businesses \cite{2}. With the addition of mask mandates and COVID-19 vaccines, American communities have been able to return to more normal operations. As lockdown measures lifted, universities across the United States handled reopening in a number of ways. Some schools opted to lift all restrictions, while some maintained virtual classes; some institutions use a mixed approach, holding classes both online and in person. 
	
Professors holding virtual classes have approached lectures in several ways. One approach has been to record video lectures that students can watch whenever they choose. This asynchronous modality offers flexibility for the student, but lacks social interaction between students and professors. Another approach makes use of video call platforms, such as Zoom, which provides features like screen-sharing, whiteboards, polls, and chat rooms \cite{3}. This online modality does not offer the same scheduling flexibility; however, it does provide students a way to safely interact with each other and their professors. While not the same as in-person instruction, synchronous video instruction is regarded by many as a good substitute. The mixed-modality approach is a combination of the in-person, asynchronous, and online modalities. This usually involves video calls or prerecorded lectures for part of the week, along with small in-person classes for the other part. In addition to the varying modalities, students have been experiencing variations in workload, available resources, and support from faculty. One of the questions this paper will examine is: which modality appears to have the best learning outcome for students?
	
COVID-19 has impacted not only students’ educational experiences at universities, but residential life. With restrictions on gathering sizes, reduced capacity in dining facilities, and limited access to sporting events, many of the ways in which students interact have altered. Depending on COVID-19 case counts within a community, the availability of study spaces could be limited and some on-campus facilities might be closed completely. Additionally, students worry about their campuses shutting down again and about the health risk of living in colleges’ residential communities. In the wake of these restrictions and concerns, college students have experienced an increase in negative emotions in relation to the pandemic \cite{4}. This has led to concerns about the wellbeing of students enrolled in higher education. This paper intends to examine several questions. First, how has COVID-19 and changes in teaching modality affected students’ mental health? Secondly, what factors appear to play the most significant role in causing a decline in students’ mental health? 

Several studies have sought to document the concerns and worsening mental health of college students in the United States. Son et al. \cite{Scot1} conducted a survey and statistical analysis to assess the impacts of the pandemic on students. Specifically, they use descriptive statistics, thematic analysis, and PSS scores to examine the results of their survey. Wang et al. \cite{Scot2} follow a similar approach to investigate the intensity of anxiety and depression among college students. They use PHQ-9 and GAD-7 to measure depression and anxiety, respectively. They then used ANOVA to find significant relationships between factors. Like Son et al. \cite{Scot1} used thematic analysis as part of their research. Lin et al. \cite{Scot25} performed a statistical analysis of data focusing on the mental health of Chinese international students in the United States. These articles primarily focus on identifying rates of depression and anxiety among college students and identifying their concerns and stressors in relation to factors such as COVID-19, living environment, financial conditions, and academic work. These findings show that rates of depression, anxiety, and stress increased during the pandemic. This demonstrated increase in poor mental health among college students led to our own research to identify what factors might influence students’ mental health. 

Alharthi \cite{Scot3} uses artificial intelligence to build a model predicting a student’s level of anxiety. They collect general information and information on stressors related to the pandemic, finding that AdaBoost and neural networks are the best performing models based on area under the curve (AUC) and that a neural network has the highest accuracy. Additionally, this research identifies the top predictors of anxiety as gender, a support system, and family fixed income. Wang et al. \cite{WangDonna} identifies several other features that explain variance in coping, based on survey results subjected to a hierarchical multiple regression analysis. This analysis finds that “age, support of family and friends, support of children’s school, use of alcohol and substances, level of trust/satisfaction with national government, being overwhelmed by the amount of COVID-related information, and level of life disruption” account for roughly 12 percent of variance in levels of coping. The results of these papers align with our own analysis, although our top predictors vary. 

Herbert et al. \cite{Scot5} conducted an extensive survey among college students in Egypt and Germany, with the goal of investigating mental health, behavior, and subjective experiences during the pandemic. Along with a statistical analysis including descriptive statistics and a correlation analysis, Herbert et al. \cite{Scot5} use two machine learning regression models: Gradient Boosting Regression and Support Vector Regression. They also apply linguistic analysis to some of their data. Their results indicate that students are experiencing above-average levels of anxiety and depression. They also found that the majority of students reported weight gain, higher food intake, and less exercise than before the pandemic. This study also found that the majority of students struggle with “self-regulated learning” \cite{Scot5}. Based on results from the machine learning models, it appears that personality and subjective experience are related in ways not shown in the statistical analysis.  

Studies by Ren et al. \cite{Scot6} and Khattar et al. \cite{Scot7} also use machine learning to analyze the effects of the pandemic on students’ mental health. Ren et al. \cite{Scot6} look for the factors behind depression and anxiety during the pandemic amongst college students in China. Using the Akaike Information Criterion and multivariate logistic regression, they find several factors influencing students’ mental health, including exercise frequency, use of alcohol, isolation, and the effect of the pandemic on their families’ economic statuses. Khattar et al. studies the effects of the pandemic on students in India. Their statistical analysis shows that COVID-19 lockdowns have negatively affected the social lives of students and that online learning does not provide a sense of connection, structure, or adequate instruction time. 

Studies such as that of Al-Mawee et al. \cite{Al-Mawee} focus on the efficacy of online learning for college students. A survey was conducted at the University of Michigan to determine how students felt about online learning, and the results indicate both positive and negative features of this modality. These results indicate that the year of schooling (i.e., freshman, sophomore) and type of college significantly impact a student’s perception of online learning, with new students typically having a more negative experience. Students seem most concerned about lack of interaction with instructors and peers and with feeling that they learn less in a virtual setting. Positives of online learning include flexibility of location, time, attendance, and assignments. Other studies, including (Dhawan et al. \cite{Scot9}, Gonzalez-Ramirez et al. \cite{Scot10}, Mukhtar et al. \cite{3}, and Zeng et al. \cite{Al-Mawee}), also seek to understand students’ and also instructors’ perspectives on distance learning and aim to identify the strengths and weaknesses of online learning. 

Machine learning research has also investigated and predicted mental health outcomes in relation to COVID-19 across other demographics. These include healthcare workers (Ćosić et al. \cite{Scot13}, Dolev et al. \cite{Scot14}, Rezapour et al. \cite{rezapour2021machine}, and Wang et al. \cite{Scot16}), adolescents and children (Viner et al. \cite{Scot17}, and Ntakolia et al. \cite{Scot18}) and older adults in assisted/community living (Vahia et al. \cite{Scot19}). Some studies have used machine learning to examine social media use and its relationship with mental health during the pandemic (Valdez et al. \cite{Scot20}, Xue et al. \cite{Scot21}, Low et al. \cite{Scot22}, Guntuku et al. \cite{Scot23}, and Ryu et al. \cite{Scot24}). 

In this paper, we use large-scale online surveys across the world, entitled ``Impact of the COVID-19 Pandemic on Life of Higher Education Students” \cite{data}. We select Question 25a as our target variable. Question 25a asks respondents to indicate how often they feel joyful while attending, studying for, or preparing for classes since the outbreak of COVID-19 on a scale of 1 to 5. On this scale, 1 is never, 2 is rarely, 3 is sometimes, 4 is often, and 5 is always. We then examine Question 25d to obtain higher accuracy models. Question 25d asks students to rate how often they feel frustrated while attending, studying for, or preparing for classes since the outbreak of COVID-19 on the same scale. We view joy and frustration as indicators of students’ emotional wellbeing. Emotional wellbeing is defined as ``happiness, confidence and not feeling depressed” \cite{Coverdale}, and is a mental health symptom \cite{Keyes}. Students who feel less joy and more frustration (students with worse emotional wellbeing) likely have worse mental health than those who feel joy more often and frustration less often. We shift from examining Question 25a as the target variable to Question 25d because they appear to be opposites. Many of the same factors appear in our models for both, and, as makes sense intuitively, variables that have positive correlations with joy tend to have negative correlations with frustration. Additionally, Question 25d seems a better target variable than Question 25a because it has stronger correlations with the other variables. We are interested in determining which factors appear to be most common among students who report worse emotional wellbeing, because these may also be the factors that have negative impacts on mental health. 

Our study contributes to the existing literature on the subject of student mental health during the COVID-19 pandemic in several ways. First, to our knowledge this is the first study where machine learning is used to analyze mental health data on college students from the United States. Second, the data we analyze covers academics, financial situations, perceptions of the pandemic, and emotional wellbeing. Specifically, regarding academics, the data and our analysis delve deeply into the modalities of online learning and their impact on students’ mental health. Our paper also extensively discusses the models we constructed and shows results from several machine learning algorithms, while many papers on this subject focus on a single model. 

The remainder of this paper is structured as follows: The following section, \nameref{Section2}, provides an artificial intelligence review. Section \nameref{Section3} discusses the methodology, describes the experimental framework used to find the top predictors of students’ emotional wellbeing, and presents computational results. Section \nameref{Section4} discusses and analyzes the top predictors of students’ emotional wellbeing obtained by machine learning methods. Section \nameref{Section5} includes the most important results and observations. Section \nameref{Section6} summarizes our overall findings.

\section*{Artificial Intelligence Review}\label{Section2}
Artificial intelligence (AI) is a field combining computer science and robust datasets to build machines capable of performing human-like tasks. AI enables machines to learn from experience and adjust to new inputs \cite {Winston}. Machine learning (ML) is a subfield of artificial intelligence that focuses on the use of data and algorithms and automates analytical model building. AI offers effective means of responding to the global public health emergency caused by COVID-19, such as detecting higher risk among patients and potential drugs for COVID-19 \cite{Fang,Ong}. Several supervised and unsupervised methods have been employed to analyze the impacts of the COVID-19 pandemic on the mental health of frontline workers \cite{rezapour2021machine,rezapour2021hidden}, and to examine the relationship between underlying medical conditions and COVID-19 susceptibility \cite{rezapour2021machineC}. Unsupervised learning is a type of machine learning in which an algorithm is used to analyze and cluster unlabeled datasets. Unsupervised machine learning and statistical learning algorithms can be used to determine the relationship between variables in a dataset. Supervised learning algorithms can learn and generalize from historical data to make predictions about new data. In supervised learning, we convert the learning problem into an optimization problem by defining a loss function as the objective function. Optimization methods (e.g., \cite{1c,2c,3c,4c,5c,6c,7c,8c,9c,10c}), which are often used to minimize a loss or error function in the model training process, play an important role in the speed-accuracy trade-off of machine learning algorithms. In the remainder of this section, we briefly review all supervised and unsupervised learning methods used in this paper.

\subsection*{Data Cleaning}
Data preprocessing in machine learning is an important step that helps to improve the quality of data so that meaningful insights can be extracted. Data preprocessing includes data cleaning, normalization, transformation, feature extraction and selection, etc. \cite{Kotsiantis,Salvador}. In this paper, our target dataset, which is the data collected by the Faculty of Public Administration at the University of Ljubljana, Slovenia in conjunction with an international consortium of universities, is categorical and encoded. The normalization, transformation, and feature extraction process are therefore unnecessary; however, it is vital to identify and correctly address missing values in the data before employing any ML model. This helps to prevent inaccurate conclusions and inferences from the data. Basically, there are two ways to handle missing categorical data: the deletion of rows with missing values and imputation. 

A simple conventional method of handling missing values and preparing data for ML analysis is to delete all rows containing missing values. Python’s pandas library provides a function to remove all such rows or columns \cite{McKinney}. While the complete removal of data with missing values might result in a robust and trustworthy model, it is unwise to delete instances containing missing values because instances with missing values also contain meaningful information. We can use ``the deletion of rows with missing values'' as an initial or alternative method for dealing with missing entries. 

The most common approaches for dealing with missing features involve imputation, the process of replacing missing data with substituted values \cite{Hastie}. Multiple imputation is a general approach to the problem of missing data, and aims to allow for uncertainty about the missing data by creating several different plausible imputed data sets and combining results from each of them \cite{Van,SaarTsechansky,Liu}. The $k$-Nearest Neighbors algorithm (KNN) classifies a new observation by a majority vote of its neighbors and does not have the training phase. KNN calculates the distance from the new observation point to all seen data points, and then the new observation class label is assigned to be the class that is the most common among its $k$ nearest neighbors \cite{Bishop}. KNNImputer by scikit-learn, a widely used method for imputing missing values, is based on KNN algorithm \cite{Pedregosa,Troyanskaya}. 

Before employing any ML model, we must determine whether the data is imbalanced. If a data set is imbalanced, which occurs when the classification categories are not approximately equally represented, inferences could be inaccurate. Nitesh Chawla, et al. \cite{Chawla} proposed a technique called the Synthetic Minority Oversampling Technique (SMOTE), which synthesizes new examples of the minority classes. The SMOTE is a combination of over-sampling the minority classes and under-sampling the majority classes to achieve a better classifier performance. The SMOTE algorithm draws a random sample from the minority class, identifies the k nearest neighbors, takes one of those neighbors and identifies the vector between the current data point and the selected neighbor, multiplies the vector by an appropriate random number, and finally adds this to the current data point \cite{Chawla}.

\subsection*{Supervised Learning Models}
Supervised learning refers to a class of algorithms that are trained on input labeled data to predict the output correctly. Caruana et al. \cite{Caruana} provides large-scale empirical comparison among ten supervised learning methods: SVMs, neural nets, logistic regression, naive Bayes, memory-based learning, random forests, decision trees, bagged trees, boosted trees, and boosted stumps. For all supervised machine learning models, we first split the data into two subsets, a training set and a test set. We train the model on the training set while the test set is held back from the algorithm. After finding the optimal parameters of the model on the training set, we apply the trained model to the test set to determine how well the model performs on unseen data points. Below, we briefly review some of the supervised models employed in this paper. 

Multinomial Logistic Regression is an extension of binary Logistic Regression for multiclass classification problems, in which the log odds of the outcomes are modeled as a linear combination of the predictor variables \cite{Dankmar}. A multiclass classification problem can be split into multiple binary classification subproblems, and then a standard logistic regression model can be fit on each subproblem. Multinomial Logistic Regression models are supported by the scikit-learn Python machine learning library. 

Support Vector Machines (SVMs) are used for classification, regression, and outlier detection. As a classifier, SVM maps data to a high-dimensional feature space so that the boundary between the categories can be defined by hyperplanes. The transformation function that is employed, the ``kernel,'' plays an important role in obtaining the best model in each case. SVM is effective in high-dimensional spaces, specifically in cases where the number of dimensions is greater than the number of samples, a condition known as ``Large-p, Small-n'' \cite{Ingo,Marti,ed,Peter}. The algorithm finds the optimal hyperplane by finding the closest points, called support vectors, to the hyperplane. It then maximizes the margin between the hyperplane and the support vectors. One-vs-one Support Vector Machines (SVM OVO) is an appropriate extension of binary SVM for multiclass classification problems. SVM OVO splits the dataset into one dataset for each class versus every other class, which means it converts a multiclass classification dataset into multiple binary classification problems \cite {Brereton}. 

Artificial Neural Networks (ANNs), often abbreviated neural networks (NNs), are computational models that consist of three types of layers: an input layer, which receives data; at least one hidden layer that processes the data; and an output layer, from which we can obtain the result out of the network. Fully connected neural networks (FCNNs) are a type of artificial neural network where the architecture is such that all the nodes in one layer are connected to the neurons in the next layer \cite{Schwing}. There are different types of neural networks, such as Feedforward Neural Network, Recurrent Neural Network, Convolutional Neural Network, etc. Each layer in a network contains neurons that pass a signal through the network. Computations occur in the neurons determining whether they fire. In these computations, a neuron combines the output of the previous layer with a set of weights to generate a weighted sum. This sum is then input into an activation function that determines whether and to what extent the signal progresses through the network. In a general feed-forward network, each neuron activation is computed as a weighted sum of its inputs from the previous layer and it is then transformed by an activation function to return the output of the neuron. The number of hidden layers and nodes of a neural network are the hyper-parameters of the model, which means we must determine them in the beginning. The parameters in a neural network that must be determined are the weights \cite{MohamadH,SchmidhuberDeep,Artificialneuralnetwork}. 

Random Forest algorithms are ensemble learning methods and were initially proposed by L. Breiman \cite{ Breiman} in 2001. Since then, Random Forest models, which are made up of many Decision Trees, have been applied to a wide range of classification and regression problems. Random Forests first build Decision Trees on different samples, and then final output is determined by Majority Voting or Averaging for classification and regression problems, respectively. Random Forest hyperparameters, which are parameters whose values are used to control the learning process and must be set before training, include the number of Decision Trees in the forest and the number of features considered by each tree when splitting a node. The best hyperparameters are usually impossible to determine ahead of time, but the optimal hyperparameters can be selected by trial-and-error-based engineering, called hyperparameter tuning. Hyperparameter tuning plays a crucial role, as hyperparameters control the overall behavior of a Random Forest \cite{Probst}. 

Gradient boosting is a machine learning technique that converts weak learners into strong learners. XGBoost \cite{34c,35c}, LightGBM \cite{36c}, and CatBoost \cite{37c} are decision-tree-based ensemble supervised learning algorithms that follow the principle of gradient boosting. The XGBoost was introduced by Chen et al. \cite{34c,35c} and is one of the most popular forms of decision-tree-based ensemble algorithms for tabular data due to its scalability, performance, and execution speed. In the learning process, XGBoost minimizes a regularized loss function and new decision trees are added one by one to improve the prediction of the previous trees in the model. The XGBoost differs from the Random Forest in that the XGBoost stops constructing the tree to a greater depth if the gain from a node is found to be minimal, while the Random Forest might overfit the data; however, Random Forests are easier to tune than Boosting algorithms and will likely not overfit the data if the data is neatly pre-processed and cleaned. 

CatBoost, introduced by Dorogush et al. \cite{37c}, is a weighted sampling version of Stochastic Gradient Boosting that uses one-hot-encoding for categorical data. CatBoost employs a technique called Minimal Variance Sampling (MVS), which is a weighted sampling version of Stochastic Gradient Boosting, and a combination of one-hot encoding and an advanced mean encoding for categorical data. CatBoost utilizes two feature importance methods. The prediction-value-change sorts features based on prediction changes if a feature value changes, and the loss-function-change sorts features based on the difference between the loss value of the model with and without a given feature. One important way in which CatBoost differs from XGBoost is that CatBoost constructs trees where the height of the left and right subtree of any given node differs by no more than one. This type of tree is called a balanced tree \cite{37ccc}.

\subsection*{Feature Selection}

A feature is an individual, measurable property of the process being observed. Feature Selection is the process of reducing the input features in a model by using only relevant features, which contribute the most to prediction, and getting rid of noise in the data \cite{Chandrashekar}. To eliminate irrelevant features, we need to set a feature selection criterion so that the relevance of each feature to the output can be measured. Supervised Feature Selection differs from unsupervised dimension reduction methods such as Principal Component Analysis (PCA) \cite {Alpaydin26} in the sense that important features can be independent of the rest of the data \cite {Law26}. In this section we briefly introduce feature selection methods using supervised learning and unsupervised learning methods. 

In the feature selection method based on Decision Tree, a fundamental splitting parameter such as Entropy, Information Gain, Gini Index, etc. is selected as the criterion, and importance scores are calculated based on the reduction in the chosen criterion. The same approach can be used for ensembles of Decision Trees, such as the Random Forest, XGBoost, LightGBM, and CatBoost \cite{Rubinstein26,Zhou26}. 

On the other hand, in the Permutation Feature Importance technique, feature importance scores are calculated based on the increase in the model’s prediction error after permuting the feature. In other words, a feature is considered ``important'' if shuffling its values increases the model error; this implies that the model relies on the feature for the prediction. The Permutation Feature Importance measurement was introduced by Breiman \cite {Breiman26} in 2001. 

After we select the most important features using supervised learning methods, the decision tree-based methods, or the Permutation Feature Importance technique, we need to understand the predictive relationship between response and predictor variables. Correlation coefficients can be used to measure how the value of two different variables varies with respect to each other. If there are multiple important features and the goal is to find correlations among all these features and store them using an appropriate data structure, a correlation heatmap can be used. A correlation heatmap is graphical representation of a 2D-correlation matrix representing correlation between features. The value of a correlation coefficient can take any values from -1 (negative correlation: when one variable increases, the other variable decreases) to 1 (positive correlation: when one variable increases, the other variable also increases). If the correlation value is zero, there is no correlation between two variables \cite{Wilkinson26}.

\section*{Methods}\label{Section3}
\subsection*{Data Resources}

To capture students’ perceptions of the impact of COVID-19, the Faculty of Public Administration at the University of Ljubljana, Slovenia, cooperating with an international consortium of universities, other higher education institutions, and students’ associations, launched one of the most comprehensive and large-scale worldwide online surveys entitled ``Impact of the COVID-19 Pandemic on Life of Higher Education Students” \cite{data}. The procedures of this survey comply with the provisions of the Declaration of Helsinki regarding research on human participants. Ethical committees of several of the higher education institutions involved approved this study, including the University of Verona (protocol number: 152951), ISPA – Instituto Universit 'ario (Ethical Clearance Number: I/035/05/2020), University of Arkansas (IRB protocol number: 2005267431), Walter Sisulu University (Ethical Clearance Number: REC/ST01/2020) and Fiji National University (CHREC ID: 252.20) \cite{data}. Aristovnik et al. \cite{data} confirm that all experiments were performed in accordance with relevant guidelines and regulations. All participants were informed about all details of the survey and provided informed consent before their participation. They provided consent to take the survey by clicking on a ``next page” button, which was stated on the first page of the online questionnaire. Participation in the survey was both voluntary and anonymous, and participants were able to leave the survey at any time without consequence. The survey was only available to participants enrolled in an institution of higher education and at least 18 years of age. No participants in this study were minors, nor was any medical information collected in the survey \cite{data}.

\subsection*{Python Codes Resources and Target Variables}
The Python code used in Subsection \ref{Computational Process} can be found in the paper’s Github repository \cite{Git}.

 \subsection*{Computational Process}\label{Computational Process}
This study aims to determine significant predictors of the emotional wellbeing of university-level students in the United States during the COVID-19 pandemic. To accomplish this goal, we target Question 25a and Question 25d (see Table \ref{Target Variables}), then train all possible machine learning models (multiclass classifiers) to find the most accurate and robust model. Finally, by examining robust machine learning models and utilizing their feature importance, we identify the top predictors of the emotional wellbeing decline among university-level students in the United States.

 \begin{table}[H]
 \begin{tabular}{ |c||p{8.5cm}|  }
 \hline
 \multicolumn{2}{|p{14cm}|}{Question 25: Please rate to what extent have you felt the following emotions while attending your classes and studying and preparing for them since the outbreak of COVID-19 in your country. } \\
 \hline
Question 25a: Joyful& 1: Never, 2: Rarely, 3: Sometimes, 4: Often, 5: Always\\
 \hline
Question 25d: Frustrated   & 1: Never, 2: Rarely, 3: Sometimes, 4: Often, 5: Always   \\
 \hline
\end{tabular}
\caption{Target Variables}
\label{Target Variables}
\end{table}
From a psychometric perspective, Question 25a and Question 25d are similar; both address emotional wellbeing. Therefore, we can treat Question 25a and Question 25d as two target dependent variables (negative correlation). We begin with Question 25a as a target variable and reduce the dimension of features (variables) by determining the most related features (top predictors) to Question 25a. To prepare the original dataset ``Final Dataset 1st Wave” (see \cite{data}) for an artificial intelligence-based analysis, we take a few steps to determine which variables are redundant or unrelated to our target variables. 

\begin{itemize} 
 \item \textbf{Final Dataset 1st Wave (FD1W)}: The original dataset collected by Aristovnik et al. \cite{data} is a 31212 by 161 (31212 participants and 161 questions) tabular dataset containing responses from students from many countries in the world with 1991207 missing values (NAN). 
  
\item \textbf{Cleaned-FD1W}: If we remove all rows that contain at least one missing value from Final Dataset 1st Wave (FD1W), we have a clean 55 by 161 tabular dataset. Therefore, there are only 55 rows (responses) with no missing values (55 participants fully answered 161 questions) from Afghanistan, Bangladesh, Bosnia and Herzegovina, Chile, China, Croatia, Ecuador, Egypt, Georgia, India, Indonesia, Iran, Japan, Kenya, Kyrgyzstan, Malaysia, Mexico, New Zealand, Oman, Pakistan, Poland, Portugal, Turkey, United States of America, and Uzbekistan. This subdataset is called \textit{Cleaned-FD1W}.

\item \textbf{US-FD1W}: If we remove all rows corresponding to all countries but the United States of America from Final Dataset 1st Wave (FD1W), we have a 392 by 161 tabular dataset with 26329 missing values (NAN). This subdataset is called \textit{US-FD1W}.

\item \textbf{US-FD1W-25(a)}: If we consider Question 25a as the target variable and remove all rows with no response for Question 25a from the US-FD1W, and keep all columns corresponding to Questions 4, 10b, 10c, 10d, 10e, 13, 14c, 15, 16a, 16d, 16e, 18a, 18c, 19b, 19e, 19g, 20b, 20c, 20e, 21b, 21c, 21e, 21h, 21i, 21j, 22e, 25a, 25b, 25c, 25d, 25e, 25f, 25g, 25h, 25i, 25j, 27, 28, 29, 32, 34, 35a, 35b, 35c, 35d, 36d, 36h, 36i, and 36j (we explain in the following subsections why these are important) and remove the remaining columns, then we have 246 rows, 49 columns and 2432 missing values. This subdataset is called \textit{US-FD1W-25(a)}.

\item \textbf{US-FD1W-25(d)}\label{US-FD1W-25(d)}: If we consider Question 25d as the target variable and remove all rows with no response for Question 25d from the US-FD1W, and keep all columns corresponding to Questions 4, 10b, 10c, 10d, 10e, 13, 14c, 15, 16a, 16d, 16e, 18a, 18c, 19b, 19e, 19g, 20b, 20c, 20e, 21b, 21c, 21e, 21h, 21i, 21j, 22e, 25a, 25b, 25c, 25d, 25e, 25f, 25g, 25h, 25i, 25j, 27, 28, 29, 32, 34, 35a, 35b, 35c, 35d, 36d, 36h, 36i, and 36j and remove the remaining columns, then we have 245 rows, 49 columns and 2405 missing values. This subdataset is called \textit{US-FD1W-25(d)}.
\end{itemize}

In Supporting Information Section, the \nameref{Data Preprocessing} discusses how the {US-FD1W-25(d)} data set is derived from the original data set {Final Dataset 1st Wave (FD1W)}. To avoid confusion, the list of obtained datasets after taking each step is given below. The \nameref{Data Description} briefly describes the variables.  

\subsubsection*{Machine learning analysis of \textit{US-FD1W-25(d)} with 5 classes}\label{Section 3.2.5}
\begin{figure}[H]
\caption{Distribution of examples among classes of Question 25d in the \textit{US-FD1W-25(d)} dataset after KNN imputation method is applied}
\centering
\includegraphics[width=.5\textwidth]{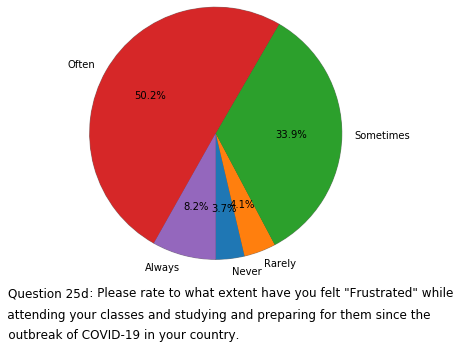}
\label{fig:f5}
\end{figure}

In this subsection, we apply multiple machine learning models to the \textit{US-FD1W-25(d)} dataset, considering Question 25d the target variable. To keep important information and treat missing values appropriately, we use KNN-imputer method again, and the result is a clean (with no missing values) 245 by 49 tabular dataset. However, the distribution of examples among classes of Question 25d is not even (see Figure \ref{fig:f5}). Hence, for each model, we split the data into training (75\%) and test (25\%) sets, then apply SMOTE on the training set, and finally test the model on the test data set. 

\begin{figure}[H]
\caption{Random Tree accuracy scores for multiple number of trees with a maximum depth of 20 on the \textit{US-FD1W-25(d)} dataset before and after SMOTE is applied.}
\centering
\includegraphics[width=\textwidth]{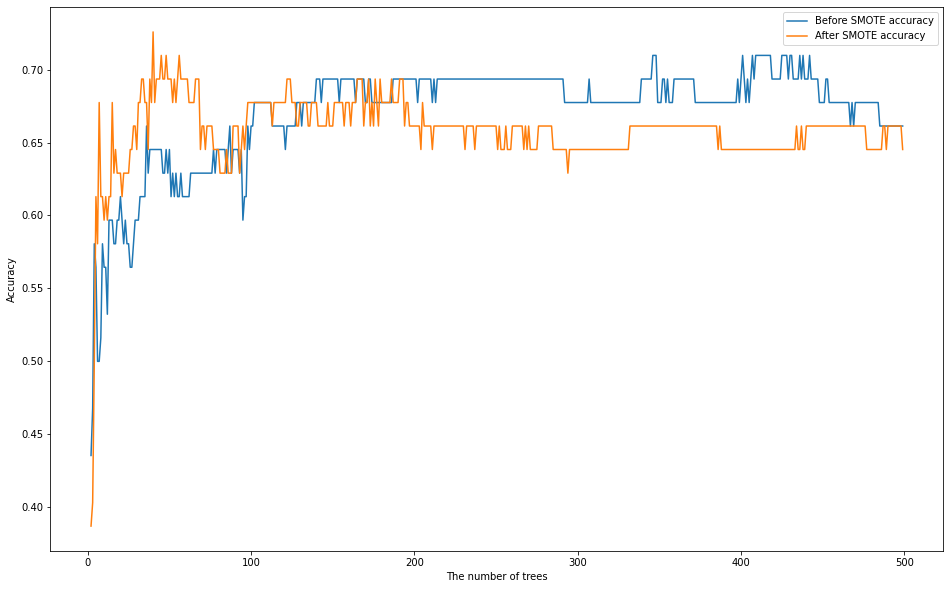}
\label{fig:re1}
\end{figure}

We first train a Random Forest with multiple number of trees with maximum depth of 20. Figure \ref{fig:re1} displays their test accuracies before and after SMOTE is applied. Table \ref {table:re1} displays the maximum, minimum, and average of accuracy scores of Random Forests before and after SMOTE is applied.  

\begin{table}[h!]
\begin{center}
\begin{tabular}{||c c c c||} 
 \hline
 & Maximum & Minimum & Average of \\ 
 & accuracy score& accuracy score & accuracy scores \\ [0.5ex] 
 \hline\hline
Before SMOTE & 0.7097 & 0.4355 & 0.6708 \\
 \hline
After SMOTE& 0.7258 & 0.3871 & 0.6583\\ [1ex] 
 \hline
\end{tabular}
\caption{The maximum, minimum and the average of accuracy scores of Random Forests where the number of trees is tuned between 2 and 500 before and after SMOTE is applied.}
\label{table:re1}
\end{center}
\end{table}

\begin{figure}[H]
\caption{SHAP values of a Random Forest with 364 trees with maximum depth of 20 before SMOTE is applied. Class 0: Never, Class 1: Rarely, Class 2: Sometimes, Class 3: Often, Class 4: Always.}
\centering
\includegraphics[width=.5\textwidth]{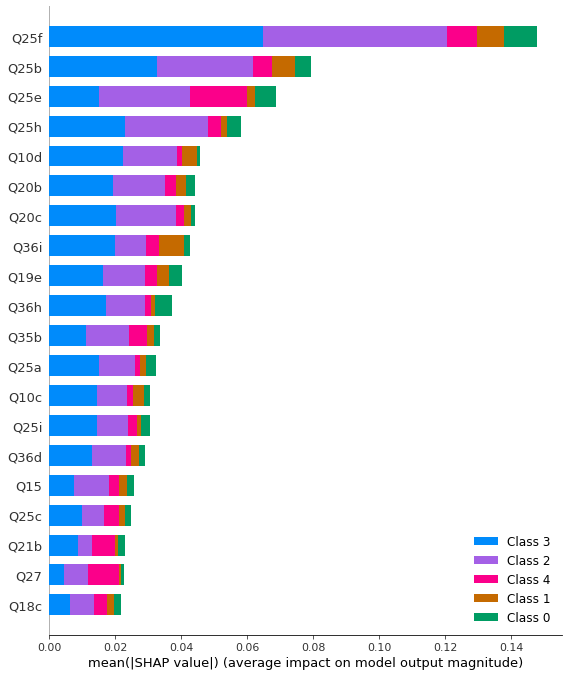}
\label{fig:re2}
\end{figure}

Before SMOTE is applied, the maximum accuracy is obtained with 346 trees with maximum depth of 20. Figure \ref{fig:re2} displays the SHAP values and feature importance of the model. 

\begin{figure}[H]
\caption{SHAP values of a Random Forest with 40 trees with maximum depth of 20 after SMOTE is applied. Class 0: Never, Class 1: Rarely, Class 2: Sometimes, Class 3: Often, Class 4: Always.}
\centering
\includegraphics[width=.5\textwidth]{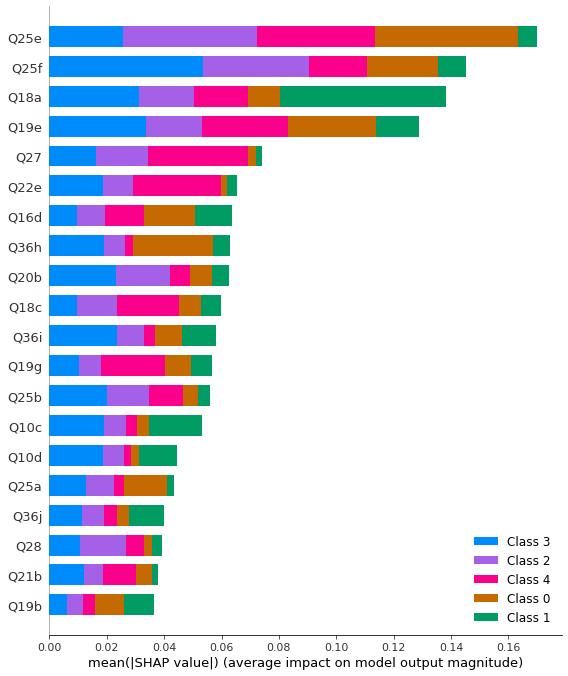}
\label{fig:re3}
\end{figure}

\begin{figure}[H]
\caption{Hyperparameter tuning:  Random Forest accuracy scores for multiple number of trees on the \textit{US-FD1W-25(d)} dataset after KNN imputation method and SMOTE are applied.}
\centering
\includegraphics[width=.8\textwidth]{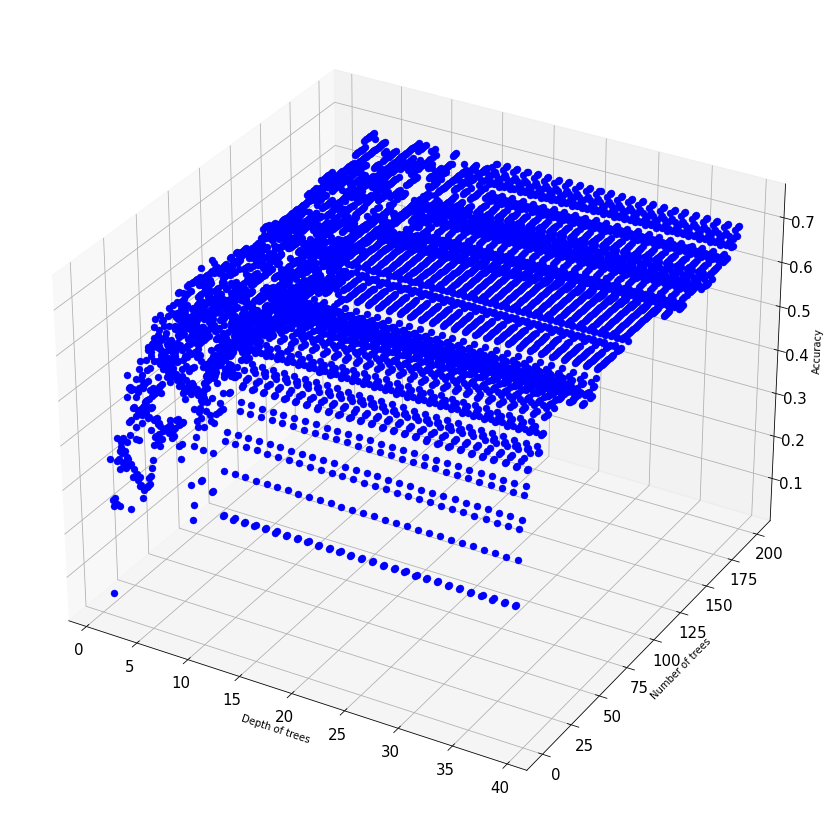}
\label{fig:ref4}
\end{figure}

After SMOTE is applied, the maximum accuracy is obtained with 40 trees with maximum depth of 20. Figure \ref{fig:re3} displays the SHAP values and feature importance of the model. 
Figure \ref{fig:ref4} displays accuracy scores of Random Forest models for multiple numbers of trees and depths after SMOTE is applied.

\begin{figure}[H]
\caption{Hyperparameter tuning:  XGBoost accuracy scores for multiple number of trees on the \textit{US-FD1W-25(d)} dataset after KNN imputation method and SMOTE are applied.}
\centering
\includegraphics[width=.8\textwidth]{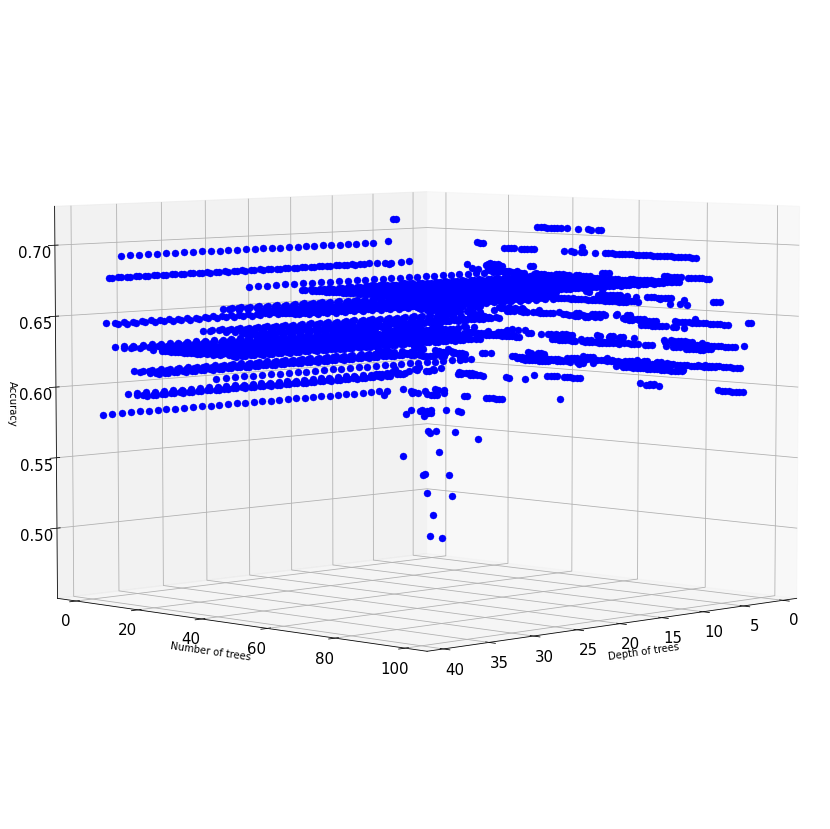}
\label{fig:ref6}
\end{figure}

\begin{figure}[H]
\caption{SHAP values of a XGBoost with 178 trees with maximum depth of 8 after SMOTE is applied. Class 0: Never, Class 1: Rarely, Class 2: Sometimes, Class 3: Often, Class 4: Always.}
\centering
\includegraphics[width=.5\textwidth]{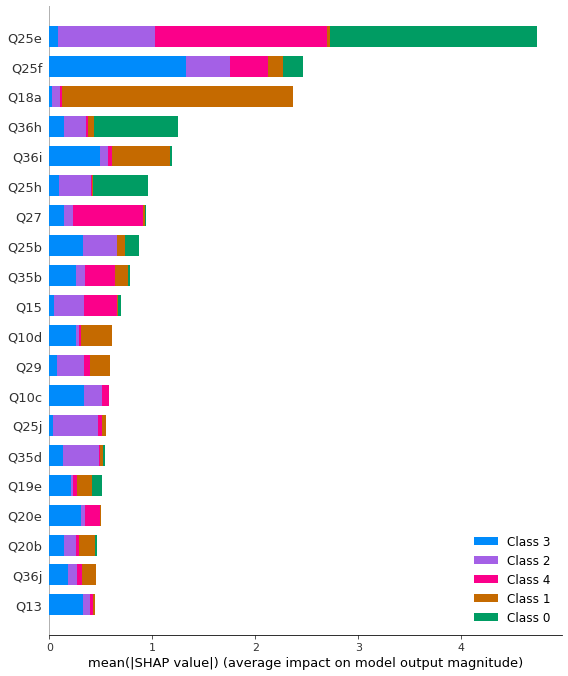}
\label{fig:ref7}
\end{figure}

We then train XGBoost models with multiple number of trees and depths. Figure \ref{fig:ref6} displays accuracy scores of XGBoost models for multiple numbers of trees and depths after SMOTE is applied. It turns out the maximum accuracy of XGBoost is 72.58\% and it is obtained with 178 trees of maximum depth of 8. Figure \ref{fig:ref7} displays the SHAP values and feature importance of the model. 

\begin{figure}[H]
\caption{CatBoost accuracy scores for multiple number of trees with a maximum depth of 8 on the \textit{US-FD1W-25(d)} dataset before and after SMOTE is applied.}
\centering
\includegraphics[width=\textwidth]{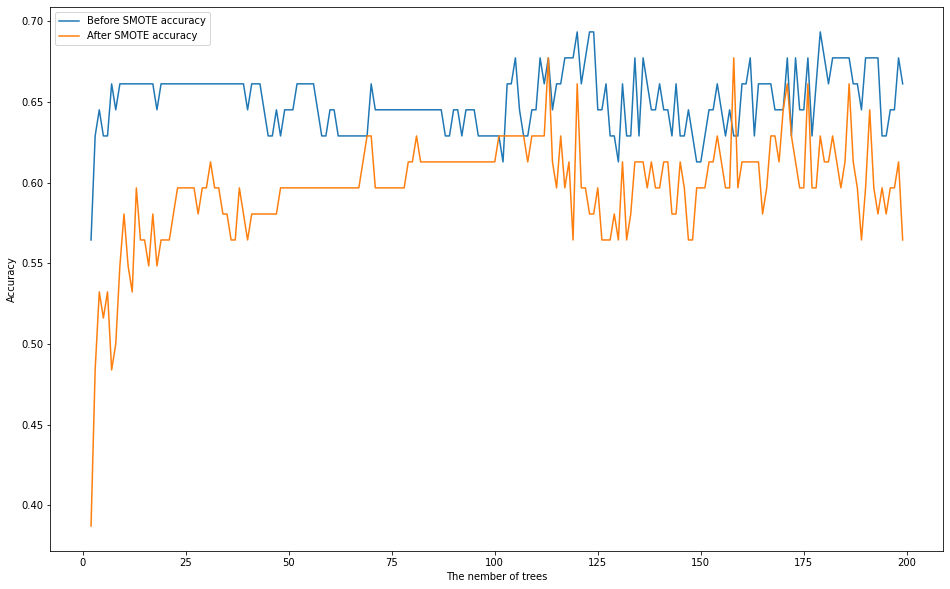}
\label{fig:ref8}
\end{figure}

Next, we train CatBoost models with multiple number of trees of maximum depth of 8. Figure \ref{fig:ref8} displays accuracy scores of CatBoost models for multiple numbers of trees after SMOTE is applied. It turns out that the maximum accuracy scores for CatBoost models before and after SOMTE is applied is below 70\%.

\begin{figure}[H]
\caption{LightGBM accuracy scores for multiple number of trees with a maximum depth of 20 on the \textit{US-FD1W-25(d)} dataset before and after SMOTE is applied.}
\centering
\includegraphics[width=\textwidth]{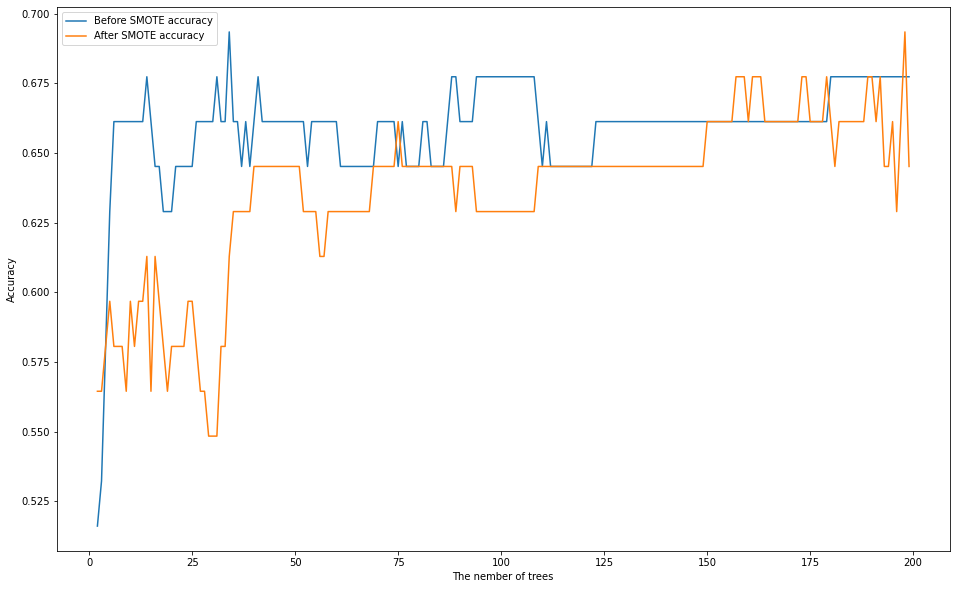}
\label{fig:ref9}
\end{figure}

We also train LightGBM models with multiple number of trees of maximum depth of 20. Figure \ref{fig:ref9} displays accuracy scores of LightGBM models for multiple numbers of trees after SMOTE is applied. It turns out that the maximum accuracy scores for LightGBM models before and after SOMTE is applied is below 70\% too.

\begin{figure}[H]
\caption{Accuracy scores of Multinomial Logistic Regression, SVM, and Neural Network with different number of hidden layers, Random Forest, XGBoost, CatBoost and LightGBM before and after SMOTE is applied for the multiclass classification problem with 5 classes.}
\centering
\includegraphics[width=\textwidth]{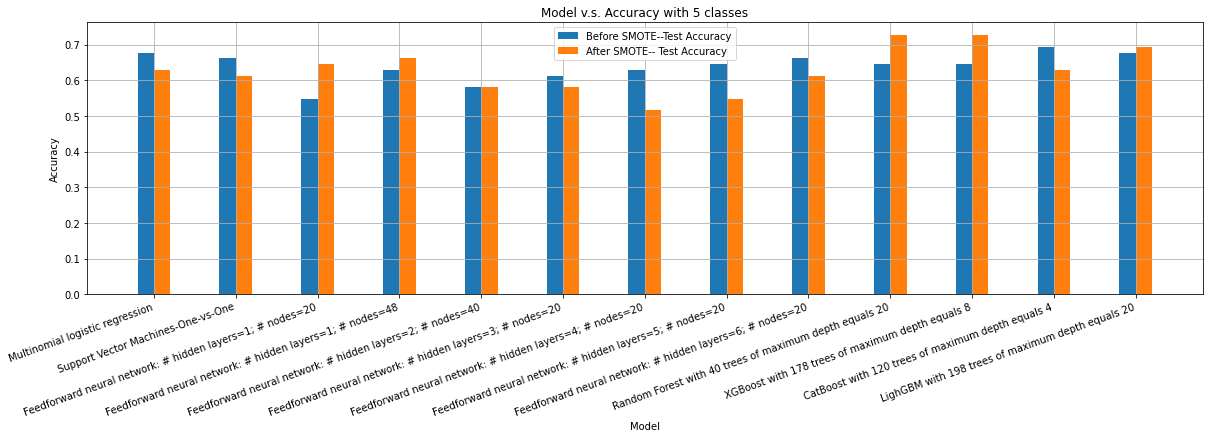}
\label{fig:ref10}
\end{figure}

Finally, we train Multinomial Logistic Regression, Support Vector Machine, and Deep Neural Network with several different number of layers and nodes. Figure \ref{fig:ref10} displays the accuracy scores of these models before and after SMOTE is applied.
As Figure \ref{fig:ref10} implies, the best obtained accuracy belongs to a Random Forest with 40 trees of a maximum depth equals 20 after SMOTE is applied. For a better understanding of the model, Table \ref{fig:ref11} displays precision, recall, and f1-score.

\begin{table}[h!]
\centering
\begin{tabular}{|c| c c c|}
 \hline
& Precision & Recall & f1-score  \\ [0.5ex] 
 \hline\hline
Accuracy&  & & 0.71 \\ 
 \hline
Macro avg & 0.47 & 0.40& 0.41 \\
 \hline
Weighted avg & 0.70 & 0.71& 0.69\\ [1ex] 
 \hline
\end{tabular}
\caption{Precision, recall and f1-score of a Random Forest with 40 trees of a maximum depth equals 20 after SMOTE is applied on the training set}
\label{fig:ref11}
\end{table}

\subsubsection*{Machine learning analysis of \textit{US-FD1W-25(d)} with 3 classes}
\begin{figure}[H]
\caption{Accuracy scores of Multinomial Logistic Regression, SVM, and Neural Network with different
number of hidden layers, Random Forest, XGBoost, CatBoost and LightGBM before and after SMOTE is
applied on the training set for the multiclass classification problem with 3 classes.}
\centering
\includegraphics[width=\textwidth]{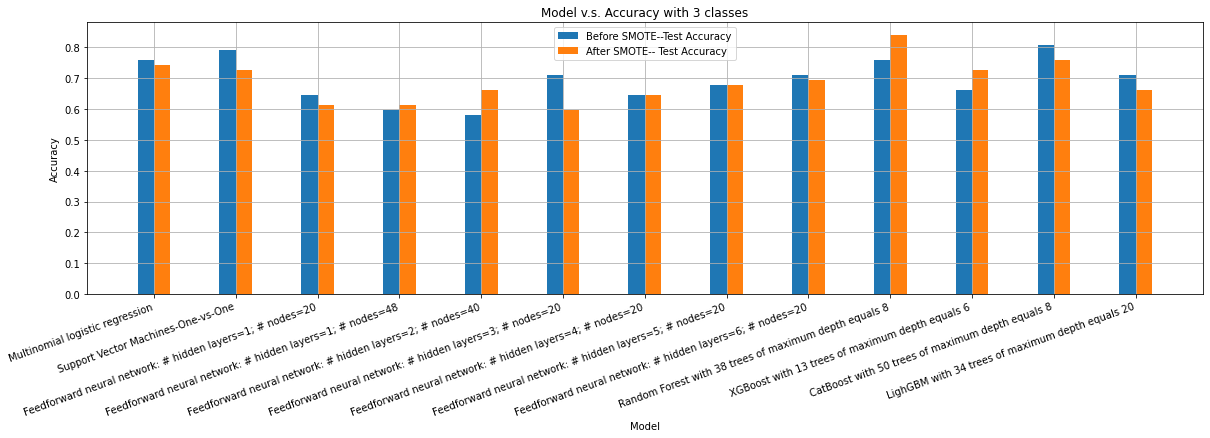}
\label{fig:part2-1}
\end{figure}

\begin{figure}[H]
\caption{SHAP values of a XGBoost with 38 trees with maximum depth of 8 after SMOTE is applied on the training set. Class 0: Never, Rarely, Class 1: Sometimes, Class 2: Often, Always.}
\centering
\includegraphics[width=.5\textwidth]{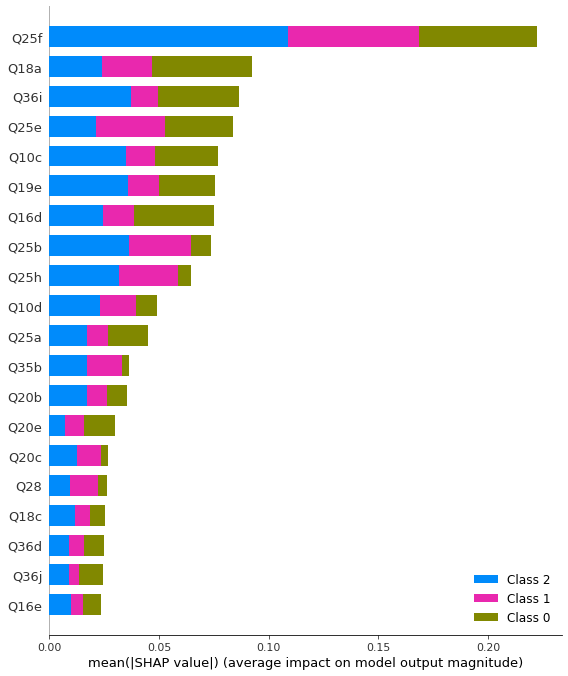}
\label{fig:part2-2}
\end{figure}

Let us examine accuracy of the aforementioned classifiers, combining instances of some classes to create a new class. We now consider three classes for Question 25(d) as below. 
\begin{table}[h!]
\begin{center}
\begin{tabular}{||c|| c c c||} 
 \hline
Class & Class 0& Class 1& Class 2 \\ [0.5ex] 
 \hline\hline
 Answers & Never, Rarely& Sometimes & Often, Always \\
 \hline
\end{tabular}
\end{center}
\end{table}
We then train all models that have used in Subsection \ref{Section 3.2.5} before and after SMOTE is applied. Figure \ref{fig:part2-1} displays accuracy scores of models before and after SMOTE is applied. As Figure \ref{fig:part2-1} implies, the best obtained accuracy, 83.87\%, belongs to a Random Forest with 38 trees of a maximum depth equals 8 after SMOTE is applied. Figure \ref{fig:part2-2} displays the SHAP values and feature importance of the model. 

\subsubsection*{Machine learning analysis of \textit{US-FD1W-25(d)} with 2 classes}
We now examine accuracy of the aforementioned classifiers where we combine instances create only two new classes. We now consider two classes for Question 25(d) as follows:
 
\begin{table}[h!]
\begin{center}
\begin{tabular}{||c|| c c ||} 
 \hline
Class & Class 0& Class 1\\ [0.5ex] 
 \hline\hline
 Answers & Never, Rarely& Sometimes, Often, Always \\
 \hline
\end{tabular}
\end{center}
\end{table}

\begin{figure}[H]
\caption{SHAP values of a Random Forest with 9 trees with maximum depth of 10 before SMOTE is applied on the training set. Class 0: Never, Rarely, Class 2: Sometimes, Often, Always.}
\centering
\includegraphics[width=.5\textwidth]{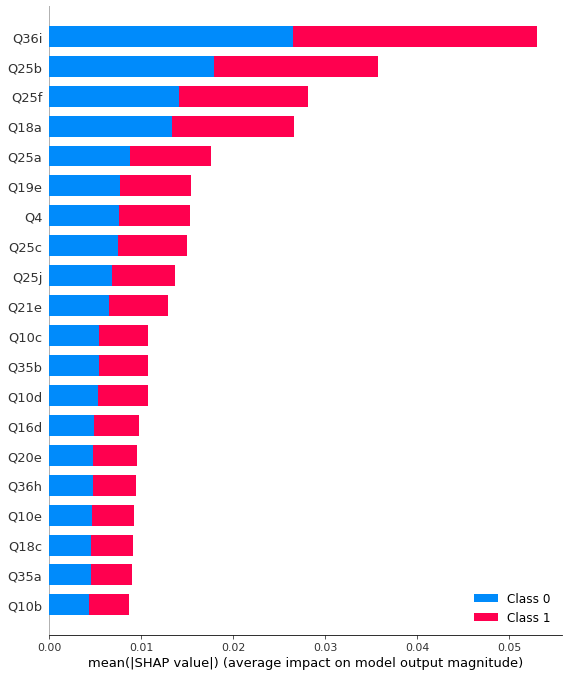}
\label{fig:part3-1}
\end{figure}

\begin{figure}[H]
\caption{SHAP values of a Random Forest with 9 trees with maximum depth of 10 before SMOTE is applied on the training set. Class 0: Never, Rarely, Class 2: Sometimes, Often, Always.}
\centering
\includegraphics[width=.5\textwidth]{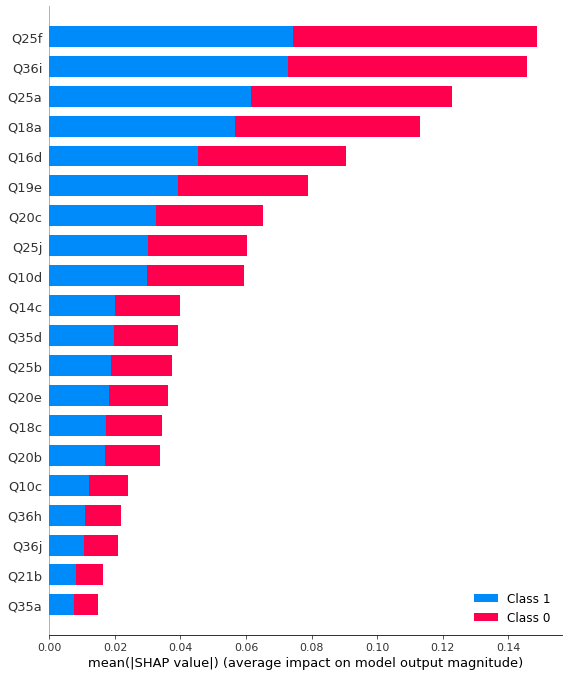}
\label{fig:part3-2}
\end{figure}

\begin{figure}[H]
\caption{SHAP values for the impact of features on model output.}
\centering
\includegraphics[width=.7\textwidth]{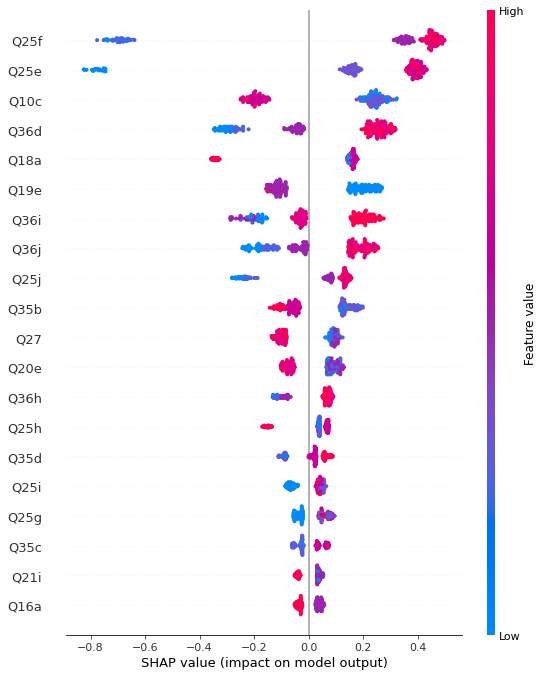}
\label{fig:part3-4}
\end{figure}

We then train Random Forest models with multiple number of trees and depths. It turns out that the best obtained accuracy for Random Forests Before and after SMOTE is applied are 95.16\% and 96.77\%, respectively. Figures \ref{fig:part3-1} and \ref{fig:part3-2} display the SHAP values and feature importance of the model before and after SMOTE is applied, respectively. Table \ref{fig:part3-3} displays precision, recall, and f1-score. Figure \ref{fig:part3-4} SHAP values for the impact of features on model output. 

\begin{table}[h!]
\centering
\begin{tabular}{|c| c c c|}
 \hline
& Precision & Recall & f1-score  \\ [0.5ex] 
 \hline\hline
Accuracy&  & & 0.95 \\ 
 \hline
Macro avg & 0.48 & 0.50& 0.49 \\
 \hline
Weighted avg & 0.91 & 0.95& 0.93\\ [1ex] 
 \hline
\end{tabular}
\caption{Precision, recall and f1-score of a Random Forest with 9 trees of a maximum depth equals 10 after SMOTE is applied on the training set}
\label{fig:part3-3}
\end{table}

\section*{Discussion}\label{Section4}
Examining the top predictors from the many different models, we have identified the questions that are highly predictive of emotional wellbeing of students in higher education in the United States. As new strains of COVID-19 emerge, we may be able to learn from the past and improve the educational environment for university students in the United States and around the world. We choose not to focus on the classes of Question 25 because a strong link between emotions is expected.

Question 18a is one of the most important predictors selected by the Random Forest for all classification problems with five, three and two classes. Question 18a and Question 18c, which ask about students’ satisfaction with the organization of supervision/mentorship since on-site classes were canceled, have negative correlations (-0.31 and 0.34, respectively) with Question 25d. Both values show that students tend to feel more frustrated when they have lower levels of satisfaction with the organization of lectures and supervision/mentorship. We believe that the correlation between Questions 18a and 25d may be explained by students feeling less engaged with the structure of online lectures. Caskurlu, Secil et al. \cite{Caskurlu} states that the way a course is designed heavily influences the learning experience of students. For example, a professor simply giving a lecture or PowerPoint presentation may bore students, leading them to feel less engaged and that they are not obtaining as much value from the course. Providing some sort of variety tends to keep students engaged and interested in the course, which seems to improve their experience \cite{Caskurlu}. Perhaps if professors work to provide this variety and work to engage further with their students, then the frustration felt among students enrolled in higher education will decrease. In fact, Caskrulu et al. \cite{Caskurlu} found that students felt that having many ``delivery modes'' in online classes helped them learn and maintain motivation, which seems to support our conjecture. Caskurlu et al. \cite{Caskurlu} also obtained results that fall in line with the correlation we found between Questions 18c and 25d. They discovered that students preferred an interactive professor and benefited from receiving feedback in class. We surmise that if instructors try to interact more with their students and provide regular feedback (in other words, if they provide more mentorship), then students will feel less frustrated in class. We also note that students from this study indicated a possibility of too much student interaction in a virtual classroom. Tools such as Zoom breakout rooms and similar features on other platforms may contribute to students’ frustration \cite{Caskurlu}. Therefore, instructors should strive to balance student interaction with traditional presentations when organizing their lectures. Making all of these improvements to the organization of online lectures will hopefully have a positive impact on the emotional wellbeing of students in higher education. 

Question 16d asks students to indicate whether their instructors have been open to student suggestions about online classes and have made adjustments based on that feedback. Question 16d is selected as an important predictor by the Random Forest for all classification problems with five, three and two classes. The correlation coefficient for the relationship between Questions 16d and 25d is -0.27, which indicates that students who report instructors being open to feedback experience frustration less often. This relationship supports our findings on question 18a, which indicates that students feel less frustrated when they are satisfied with the organization of lectures. The transition to virtual learning was not easy and was made more difficult by poor internet connections, lack of technological understanding, and other complications. To make sure that we use online modalities well, we must understand not only technological capabilities but how technology can best be used. We know that course design impacts students’ learning \cite{Peter}, and from the relationships we find between question 25d and questions 20b and 20c we see that poor learning correlates with higher levels of frustration. If instructors were to listen to student feedback and adjust their courses and lectures accordingly, there might be an improvement in their students’ emotional well-being and learning outcomes. 

Question 19e is also selected as the next important predictor by the Random Forest for all classification problems with five, three and two classes. This question asks students to rate their satisfaction with their universities’ international offices since on-site classes were canceled. The importance of this feature indicates that the pandemic may impact international students domestic students unequally. This does appear to be the case, based on various barriers to entry faced by many students coming to the United States. These barriers include campus lockdowns, travel bans, and visa restrictions, as well as general health and safety concerns \cite{Mok}. The negative correlation between Questions 19e and 25d shows that students tend to be more frustrated when they are unsatisfied with the support offered by their school’s international office. This frustration may have an impact on future students’ decisions about whether they should study in the United States, or study abroad in general. Findings from a study on Mainland China and Hong Kong students’ opinions about studying abroad indicate a decline in willingness to study overseas. In fact, only 16\% of respondents in that study showed any interest in studying abroad post-pandemic \cite{Mok}. Given that prior to COVID-19 there was a steady increase in the number of Mainland students studying internationally, it does seem that the pandemic has majorly impacted international students \cite{Mok}. Considering this in context of our own findings, we believe that universities in the United States need to reevaluate how they have been serving international students, not only to improve the lives of current students, but to continue to attract new students from abroad. Another possible explanation for the association between students’ frustration and their satisfaction with international offices is their desire to study abroad. For example, prior to the pandemic in 2017 and 2018, over 11,000 US students were studying in China \cite{Martel}. At the time of the COVID-19 outbreak, 50 US schools reported having only 405 students in China, and as of February 26, 2020, 70\% of schools were evacuating their students from the country \cite{Martel}. Many students find study abroad programs appealing, so cancellations and postponements might understandably cause frustration. 

Question 20b asks students if they feel their academic performance has improved since on-site classes were canceled, and question 20c asks if their academic performance worsened since classes were canceled. Question 20c is selected as another important predictor by the Random Forest for all classification problems with five, three and two classes. The R-value for the relationship between questions 20b and 25d is -0.32, and the R-value for questions 20c and 25d is 0.24. These correlations indicate that when a student’s perception of their performance is positive, they tend to feel less frustrated, and when their perception is negative, they tend to feel more frustrated. Past studies have examined the link between academic success and mental health, and generally mental health is used to predict academic success. For example, Eisenberg et al. \cite{revised-ref1} finds that depression is a significant predictor of low GPAs and that co-occurring depression and anxiety are associated with a lower GPA. Additionally, while Blank et al. \cite{revised-ref2} does not find a correlation between mental health and academic success amongst medical students, they do find mental health to be correlated with achievement motivation, and achievement motivation to be correlated with academic success. The findings of these studies have similar implications: that improved mental health could lead to improved academic performance. While we find similar correlations to these works, our analysis indicates that academic success is a predictor of emotional well-being. So, while the established literature indicates that better mental health may lead to more academic success, we believe that it may also be the case that academic success can lead to better mental health. Addressing the issue of mental health on college campuses from both of these perspectives by improving mental health services on campus as well as working to improve instruction, and therefore increase students’ academic success, could lead to the best possible results. 

Question 36i, which asks students if they feel that emergency childcare for essential workers is important, is selected as the one of the most important predictors of emotional wellbeing of US university students by the strongest supervised machine learning model, the Random Forest for a the classification problem with two classes. The correlation between Questions 36i and 25d indicates that students who rate this as an important policy tend to feel more frustration in school. This result is surprising because there is no obvious link between being a college student and needing childcare as an essential worker. Perhaps students feel this is an important policy because it affects family members. The importance of Question 36i, as well as 36j, could also be an indication of general actions students wish the government to enact and that the government’s overall handling of the pandemic has negatively affected the emotional wellbeing of students. This is supported by the importance of Question 35a, which is selected by the Random Forest model with 26 trees of depth 10. Question 35a asks students to rate their satisfaction with the government relative to the pandemic. The negative correlation between Questions 35a and 25d shows that students who are less satisfied with the government tend to be more frustrated. The coronavirus has affected students’ lives at and outside of school, so it would make sense that policies affecting the country they live in also majorly impact their emotional wellbeing. 

Question 10d asks students to rate their satisfaction with online lectures in the form of presentations sent from the instructor. This is the sixth most important feature selected by the Random Forest with 9 trees of depth 10 and is selected by our other top models. There is a negative correlation between Questions 10d and 25d, suggesting that students who are not satisfied with the presentations tend to feel more frustrated in class. This form of lecture has the same shortcomings we identified with audio recordings (Question 10c): a lack of community and a lack of a present and engaged instructor. With this type of lecture there is once again no classroom setting and no possibility of interaction with professors during lecture. We believe that a presentation would function better as a supplement to a lecture delivered in another format. Live lectures, video recordings, and audio recordings could all be enhanced by the inclusion of a presentation such as a PowerPoint. Caskurlu et al \cite{Caskurlu} finds that this helps keep students more engaged and motivated in class, which could reduce the frustration they feel regarding their courses.

Question 10c, which asks students to rate their level of satisfaction with lectures that are online in the form of an asynchronous audio recording (not in real-time), is another top predictor selected by all of our best models. This means that asynchronous virtual classes are more important in predicting frustration among students than synchronous sessions. The negative correlation between Questions 10c and 25d means students who feel more frustration in class tend to be less satisfied with audio recording lectures. We believe this is because audio recordings lack the features that students indicate are important for online learning. These features include a “supporting sense of community”, instructors being interactive, and instructors being present \cite{Caskurlu}. Audio recordings, and asynchronous classes in general, lack these features by design. Because they do not occur in real-time, there is no classroom setting (physical or virtual) for students to interact in. There is also no interaction between students and professors in class, because everything is pre-recorded and intended to be done on students’ own time. Audio recordings may be selected as more important by our model because of the lack of visuals to accompany the lecture. Students do not see their professor with this format, which highlights the fact that there is no interaction or instructor present. We believe that asynchronous modalities should be used in conjunction with synchronous learning to create the best online environment for students. Students do indicate that they like a variety of learning resources, so including video recordings, audio recordings, or other forms of asynchronous instruction could be beneficial in enhancing synchronous sessions. This model of online learning may be the best form of instruction when in-person classes are not an option, and hopefully will have fewer negative impacts on students’ emotional wellbeing than purely asynchronous classes. 

Question 27 is selected by the Random Forest of 40 trees with depth 20 as an important feature. This question asks students if they can cover the cost of their education when considering their monthly disposable income before the pandemic. The positive correlation between Questions 27 and 25d shows that those enrolled in higher education who struggle to cover the costs of their schooling tend to feel more frustration. We believe this is due to stress caused by financial insecurity. There is a link between financial stress and a decline in mental health, and research done by Jones et al. \cite{Martel} sought to determine what factors, including financial stress, contribute the most to the decline in mental health among those enrolled in higher education. Their findings indicate that financial stress is second only to academic stress in accounting for variance in anxiety among students. This supports our conjecture because anxiety, like frustration, can be used as an indicator of poor emotional wellbeing. Additionally, Questions 25d and 25f (frustration and anxiety) have a significant positive correlation value of 0.52. It is interesting to note that factors outside of the classroom seem to affect students’ frustration in class, as shown by the relationships between Questions 27 and 25d, as well as between Questions 19e and 25d. To address the financial concerns raised by the importance of Question 27, we would recommend that universities find a way to be more flexible regarding tuition payments. For example, providing multiple payment plans to fit students’ needs might alleviate some of the stress that damages their emotional wellbeing. Some other potentially beneficial policies would be to increase the number of scholarships offered or provide more employment opportunities for students on campus to help cover the costs of education. 

Question 35b, which asks students to rate their satisfaction with their university, is selected as the second most important predictor of emotional wellbeing of US higher education students by the Random Forest with 76 trees of depth 10. Moreover, the classes of Question 35 have negative correlations with Question 25d, however only Question 35a and Question 35b have significant correlation values: -0.24 and -0.36, respectively. The strength of the correlation between Question 25d and Question 35b is particularly interesting because it shows a relationship between students’ satisfaction with how their university has handled COVID-19 and their frustration. Specifically, it shows that students tend to be more frustrated when they are less satisfied with their school. This relationship indicates that schools may need to evaluate their handling of the pandemic and how their decisions affect students. 

A study conducted at Western Michigan University, with the goal of understanding the perspective of students on distance learning, found that their students tend to prefer face-to-face learning to online modalities \cite{Al-Mawee}. Therefore, we believe that universities should enact policies to allow for face-to-face classes. According to research from Moody et al. \cite{Moody}, across all assumptions (large or small number of social connections, highly or lowly clustered population, etc.) a “rapid decline in infections happens at relatively moderate rates of masking and vaccination.” This means that masking and vaccine mandates are effective and would allow for a safe return to in-person classes. Additionally, according to a survey conducted among college students to assess their opinions on university policies, universities with masking and vaccination mandates have higher approval ratings among their student bodies than schools without these policies \cite{Trujillo}. Based on our own analysis of the data provided in the FD1W, some other actions universities should consider which may improve students’ emotional wellbeing involve ensuring the financial security of students. These could include scholarship programs or flexibility in tuition payments to make sure students can afford the costs of education during the pandemic. 

Question 36j is also selected by all our top models. It asks students if they think delaying taxes is an important support measure the government can take in dealing with COVID-19. The positive correlation between Questions 36j and 25d shows that students who feel delayed taxes would be helpful tend to feel more frustration in school. This is another indicator of financial stress impacting the emotional well-being of those enrolled in higher education; however this time it is unrelated to the university. This is an important reminder that students are impacted by factors outside of school, and that these stressors probably contribute to the mental health crisis plaguing college campuses. 

Question 36h, selected by the Random Forest with 9 trees of depth 10, the Random Forest with 50 trees of depth 20, and the XGBoost with 178 trees of depth 8, asks students to rate how important they find deferred student loan payments in context of COVID-19. Figure 22 shows that question 36h and our target variable question 25d have a positive correlation of 0.25. This indicates that people who feel deferred loans are important also tend to feel frustrated more often. This supports a study conducted by Walesmann et al. in 2015, which found that higher yearly and cumulative student loans correlate to worse psychological outcomes even after accounting for differences between students such as income, familial wealth, and educational attainment \cite{revised-ref3}. This also aligns with the results of Jones et al. \cite{Schwing}, which shows a link between financial stress and a decline in mental health. This is alarming, given that for over thirty years the cost of tuition has increased at a much higher rate than inflation \cite{revised-ref4}. Because tuition expenses have risen so much, it is now difficult for an individual to afford a US college education without some sort of financial aid, whether from scholarships, family, or student loans. Shy \cite{revised-ref4} finds that this increase in tuition seems to be related to student loan availability. Since the availability of more affordable loans makes college enrollment more widely accessible, colleges raise prices \cite{revised-ref4}. Therefore greater numbers of students face the stress of keeping up with their loan payments, and greater numbers of students are graduating with debt, a stress that remains with them well past college. 

The relationship between Questions 36h and 25d also makes sense given that other financial features have been selected as important by our models. For instance, Question 27, which asks if the students can afford the overall costs of their education when considering their disposable income before the pandemic, has a positive correlation of 0.32 with question 36h. This implies that students under financial pressure wish for deferred student loan payments, which makes sense because this would ease their financial burden. A payment pause was put in place on March 13, 2020 in the United States on eligible student loans and interest rates on those loans was set to 0\% \cite{revised-ref6}. This is set to end August 31, 2022 \cite{revised-ref6}.  As this policy was put in place before this survey was administered, it seems that it did not fully alleviate concerns, possibly because not all student loans are administered by the government and private loans were affected by this policy, or because not all government loans were eligible for the payment pause. Ultimately, we gather from this result further evidence that students subject to financial stress are more likely to suffer from poor emotional wellbeing. While this payment pause cannot last forever, perhaps there is a way to ease back into full student loan payments. For example, an assessment of eligibility for a pause extension could be conducted based on income and savings. Another option might involve lower post-pause payments or interest rates to reduce the impact of adding this stressor back into students’ lives. 

Question 15 asks students about their preferred form of online mentorship. Possible responses were ``via video-call,'' ``via voice-call,'' ``via email communication,'' ``texting on social networks (Facebook messenger, Viber, WhatsApp, WeChat, etc.),'' ``not applicable (I had no supervisions/mentorships),'' coded 1-5, respectively. If we consider these to be ranked with 1 being the most interactive and 5 being the least, we can surmise from the negative correlation (R-value of -0.13) between questions 15 and 25d that students tend to feel more frustration when having more interactive communications with whomever is mentoring them. This result is surprising, given that our analysis also shows that students tend to feel more frustrated when they feel less mentored. This could perhaps be due to Generation Z’s preference for messaging as a form of communication as opposed to voice calls. Generation Z seems to prefer texting over calling, with 63\% of individuals saying they text daily and only 39\% saying they made calls based on data collected in 2011 \cite{revised-ref8}. This aligns with our results, which indicate that email or direct messaging are forms of communication preferred by college students over making calls. This data also showed that 43\% of Generation Z prefers communicating with others online instead of in person \cite{revised-ref8}. Social media usage has become more prevalent since 2011; between that and the COVID-19 pandemic, it is possible that Generation Z has become more comfortable operating in an online space and finds more personal interaction to be stress-inducing. Video-calls and voice-calls may be considered by some too similar to in-person interaction and therefore a source of anxiety or frustration. This would explain the relationship we see in our own analysis between virtual interaction with mentors and frustration, which implies that college students prefer more impersonal interactions with their mentors online. The reasons for this are not uncovered by our study, but one possible reason could be Generation Z’s growing comfort with virtual interactions caused by growing up with the internet and the pandemic lockdowns preventing many from interacting outside of a virtual space. 

Question 21b asks students if they have access to a desk at home. This feature is identified by the Random Forest with 40 tress of depth 20. The negative correlation between Questions 21b and 25d shows that students without access to a desk at home are more likely to feel frustrated in class. This is not an unexpected result given many students had to attend classes virtually at home. The importance of this feature emphasizes that not every university-level student has access to the same resources at home. Not having a desk, a place to work, would of course be frustrating when trying to complete classes. To mitigate this issue, colleges should strive to create a safe and healthy environment so that they may keep students on campus. One of the best ways to do this is to require vaccines and masks for faculty, staff, and students. These mandates have been shown to be effective in reducing the rate of infection across all types of communities \cite{Moody}. By enacting these policies, universities can keep students on campus and ensure that they have access to the resources they need to successfully complete their education.

\section*{Limitations of Study}
The results of our study are subject to certain limitations. The data we used did not collect psychological data other than students’ self-reporting on their emotions. Information such as symptoms of anxiety and depression was not collected as part of the survey. Therefore these datapoints are not objective measures and may be influenced by the individual respondent. Additionally, there is no baseline or control for comparison, so we cannot be certain that the participants’ emotions have changed since the pandemic. While we cannot be sure based on this data that COVID-19 has worsened students’ mental health, we are able to identify which features seem to be related to students reporting feeling negative emotions more often. We also had to address missing information in the dataset. While we used several methods, such as focusing on one country, removing rows with missing data, and using only the most related features, we would likely have a better understanding of the subject if we had more complete information. A limitation of our methodology is that the selected features we used when analyzing the US data came from importance scores from analyzing the entire dataset. While this saved some time and was more computationally efficient for building future models, it does mean that any differences between important features in the United States and global data may have been missed. While our study focused heavily on emotions and their influencing factors, future studies could seek more complete and comprehensive data on students’ mental states to grasp a better understanding of how living during the pandemic and post-pandemic has affected students’ mental health. Additionally, future work strictly using data collected from the United States could provide more nuanced results. 

\section*{Results and Recommendations}\label{Section5}

Our analysis has found that the top predictors of college students’ frustration with school are Questions 18a, 16d, 19e, 20c, and 36i, as identified by the Random Forest model with 9 trees of depth 10 after SMOTE was applied. Other notable features include Questions 10c, 10d, 27, 35b, 36h, and 36j. All the most important features displayed in Tables 2 and 3 can be generally separated into three categories that seem to have a significant impact on the emotional wellbeing of students enrolled in higher education: academic life, satisfaction with impactful organizations, and financial stressors.

Of all the variables identified in our analysis, those related to course structure, teaching modality, and academic success seem to be the most important factors contributing to students’ frustration. To alleviate the academic stress being caused by virtual classes, we believe that universities should strive to return to in-person learning. The best way to do that is to enact policies to mitigate the spread of the coronavirus, such as mask mandates and vaccination requirements. Schools with these policies tend to have more approval from their student body than those without, so in enacting these policies, schools could also reduce frustration caused by students’ dissatisfaction with their institutions’ handling of the pandemic \cite{Trujillo}. Asynchronous modalities appear to be worse for students’ emotional wellbeing than synchronous learning, so in situations where face-to-face learning is not possible, using a synchronous online modality supported by asynchronous components appears to be the best way to handle virtual learning. This format provides some semblance of a community as well as opportunity for engaging with professors.

Impactful organizations are those which affect students’ lives in a significant way; such organizations include universities, governments, banks, and other entities. In relation to our work, governments and universities are the most impactful organizations. While students can take actions to mitigate the pandemic or influence their school’s or the nation’s politics, ultimately they cannot directly change these factors. Results from a survey conducted by Wang et al. [9] support our selection of these factors. This survey was conducted across the United States as well as part of Canada with the goal of
assessing how well people are coping with the pandemic. One factor identified in that survey that contributes significantly to how well people are coping is their trust/satisfaction with the national government. Specific government policies that seem to affect students’ emotional wellbeing selected by our models are emergency childcare for essential workers, deferred student loan payments, and delayed taxes. We take this as an indication that students may be unsatisfied with the government’s response to COVID-19, and that these policies are most important to them. Wang et al. \cite{WangDonna} states that the lack of a “unified message” from the federal government may contribute to peoples’ frustration and explain their dissatisfaction with the government. This claim and our conjecture are supported by the correlation between Questions 25d and 35a, which has a correlation value of -0.24. This relationship shows a trend in which students who are more frustrated are also dissatisfied with the government’s handling of the pandemic. Students’ frustration relative to these variables could also mean that
COVID-19 itself is having a negative effect on their emotional wellbeing, and that their frustration with the government is a simply a byproduct of the pandemic.

Financial stressors are the other major category impacting emotional wellbeing. These findings support the results of Jones et al. \cite{Jones}, which state that academic stress, followed by financial stress, accounts for the largest amounts of variance in anxiety among college students. The fact that deferred student loan payments and delayed taxes are two of the policies identified as the most important to students indicates they are under financial pressure due to the pandemic. Additionally, we find that students who cannot cover the total cost of their studies with their disposable income tend to have worse emotional wellbeing outcomes. To help alleviate some of the financial stress being caused by COVID-19, we recommend that universities offer flexible payment plans for their students. This would make it easier for students to cover the cost of their education and might also allow them to avoid student loans and going into debt.

\section*{Conclusion}\label{Section6}

In this paper we present our analysis of the dataset titled, “Final Dataset 1st Wave”, collected by the Faculty of Public Administration at the University of Ljubljana, Slovenia, as well as an international consortium of universities, other higher education institutions and students’ associations. We used several machine learning and statistical models to analyze this data. The most accurate of these models is the Random Forest model with 9 trees of depth 10 after SMOTE is applied. Our other top models are the Random Forest with 38 trees of depth 8, the Random Forest with 40 trees of depth 20, and the XGBoost with 178 trees of depth 8. Our results indicate that the predictors of emotional wellbeing selected by these models can be sorted into three categories: challenges with academic life, satisfaction with impactful organizations (universities, governments, etc.), and financial stressors. Based on these results, we recommend universities strive to offer face-to-face classes, and synchronous classes supported with asynchronous components to enable the healthiest learning for students. We also recommend schools work with their students to meet their financial needs: for example, by offering various tuition payment plans to reduce financial stress. We also recommend that universities strive to understand the policies students desire to see on their campuses. It is our hope that the information provided by our analysis will help to create a better experience for those enrolled in institutions of higher education as we continue to cope with COVID-19 and the possibility of future pandemics.

\nolinenumbers

% Either type in your references using
% \begin{thebibliography}{}
% Text
% \end{thebibliography}
%
% or
%
% Compile your BiBTeX database using our plos2015.bst
% style file and paste the contents of your .bbl file
% here. See http://journals.plos.org/plosone/s/latex for 

\section*{Competing interests}
Te authors declare no competing interests.

\newpage

\appendix
\section*{Supporting Information}\label{Supporting information}
\section*{Data Preprocessing.}
\label{Data Preprocessing} In this subsection, we describe how the {US-FD1W-25(d)} data set is derived from the original data set {Final Dataset 1st Wave (FD1W)}. To address missing entries in the original dataset, and its subdatasets, we use two primary methods: the deletion of rows with missing values and imputation.

\subsubsection*{The deletion of rows with missing values from the FD1W and computational results}

If we remove all rows of the FD1W that contain at least one missing value, the result is the \textit{Cleaned-FD1W}; however, the \textit{Cleaned-FD1W} is a 55 by 161 tabular dataset, which means that we must deal with a ``large-p, small-n'' problem (see \cite{Lavelle}), because the number of predictors (features or columns) is much larger than the number of samples (datapoints or rows). 

We first select Question 25a as the target variable, split the data into training and test sets (75\%-25\%), and train possible supervised multiclass classifiers on the data. We aim to find an accurate model and analyze its feature importance scores. In large-p, small-n cases, one of the most widely used machine learning methods is Support Vector Machine (SVM). Since our target variable, Question 25a, contains five classes, we apply SVM OVO, which is appropriate for multi-class classifications. Due to a severe large-p, small-n problem, the SVM OVO model obtains a test accuracy score of 52.63\%. Since the accuracy of the SVM OVO model is not satisfactory, we do not analyze the feature importance scores of the model. We also apply the XGBoost model with $k$-fold cross validation, where $k$ changes between 2 and 20. Since there are many more features than datapoints, the obtained accuracy scores for the model are not satisfactory. Figure \ref{fig:p1} displays accuracy scores of XGBoost using $k$-fold cross validation for multiple number of folds.

\begin{figure}[H]
\caption{Accuracy scores of XGBoost with k-fold cross validation on the \textit{Cleaned-FD1W}}
\centering
\includegraphics[width=0.8\textwidth]{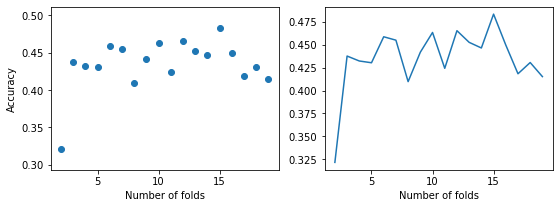}
\label{fig:p1}
\end{figure}
We then use CatBoost with multiple numbers of trees. Figure 2 displays accuracy scores of CatBoost with multiple numbers of trees. It turns out that the CatBoost with 19 trees outperforms with an accuracy score of 47.2\%; however, due to large-p, small-n related issues, the accuracy for boosting models on the \textit{Cleaned-FD1W} is not satisfactory either. To alleviate large-p, small-n related issues, where $p$=160 and $n$=55, we can reduce the feature dimension by removing unrelated features (columns) from the dataset. So we must determine which features are the least-related features for XGBoost using $k$=15 folds and CatBoost with 19 trees. Figures \ref{fig:p3} and \ref{fig:p4} display feature importance scores for XGBoost and CatBoost, respectively. 

\begin{figure}[H]
\caption{Hyperparameter tuning:  Accuracy scores of CatBoost with multiple number of trees on the \textit{Cleaned-FD1W}}
\centering
\includegraphics[width=0.8\textwidth]{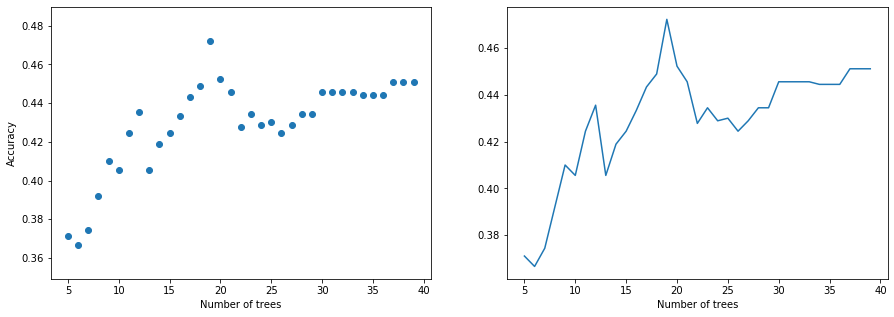}
\label{fig:p1}
\end{figure}

\begin{figure}[H]
\caption{Feature importance scores for XGBoost using k=15 cross-validation on the \textit{Cleaned-FD1W}}
\centering
\includegraphics[width=.9\textwidth]{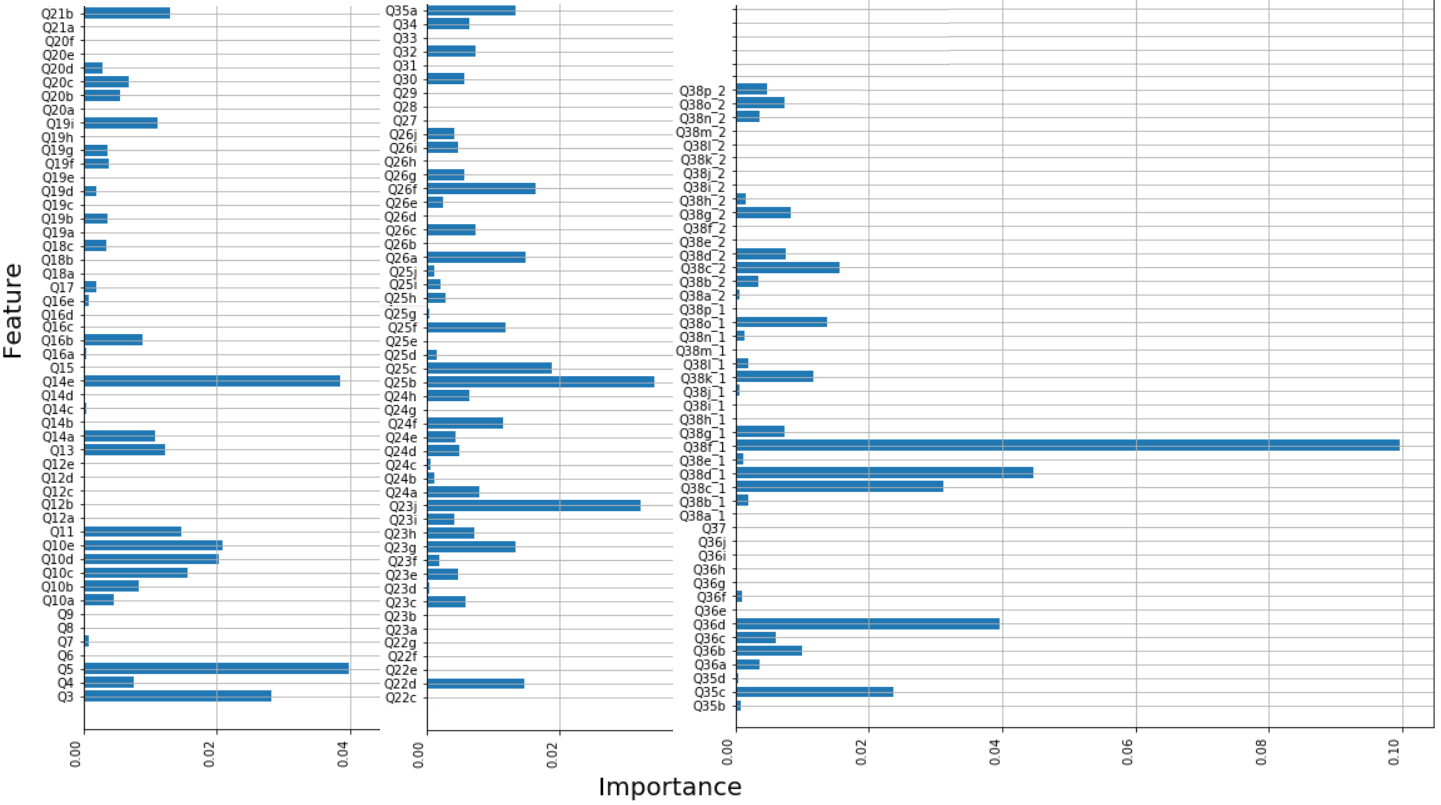}
\label{fig:p3}
\end{figure}
\begin{figure}[H]
\caption{Feature importance scores for CatBoost with 19 trees on the \textit{Cleaned-FD1W}}
\centering
\includegraphics[width=.9\textwidth]{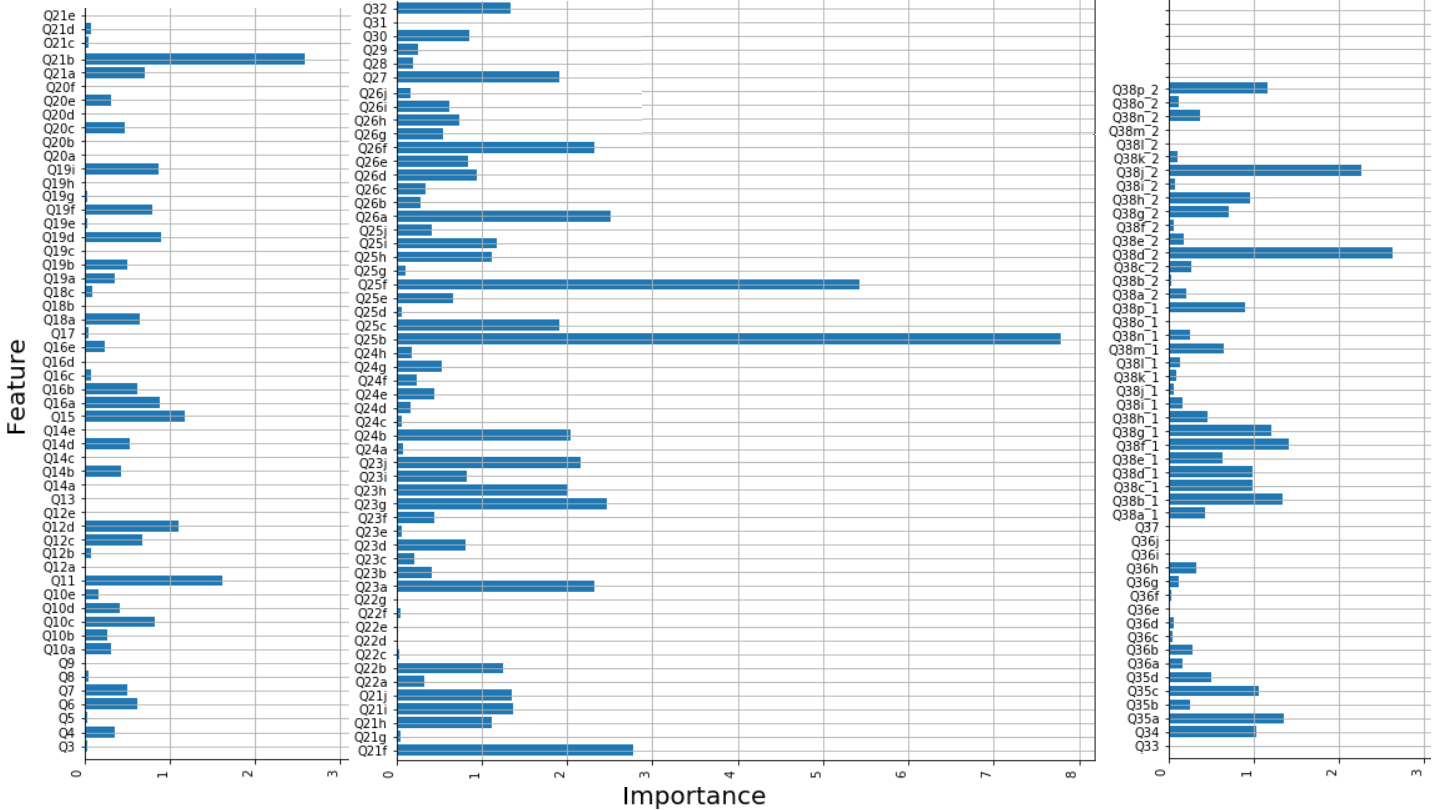}
\label{fig:p4}
\end{figure}

\begin{figure}[H]
\caption{Feature importance scores for CatBoost with 19 trees on the \textit{Cleaned-FD1W} using the related features}
\centering
\includegraphics[width=.9\textwidth]{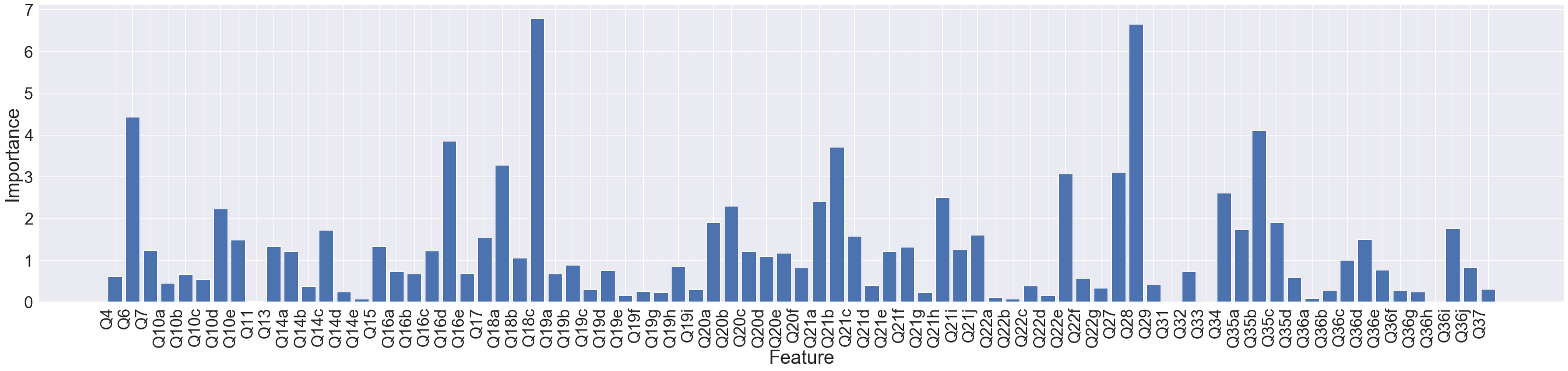}
\label{fig:p5}
\end{figure}

Using Figures \ref{fig:p3} and \ref{fig:p4}, we retain the most important features and remove unrelated features, then retrain a CatBoost with 19 trees to calculate feature importance scores (see Figure \ref{fig:p5}).

\begin{figure}[H]
\caption{Feature importance scores for CatBoost with 19 trees on the \textit{Cleaned-FD1W} using the most related features}
\centering
\includegraphics[width=.9\textwidth]{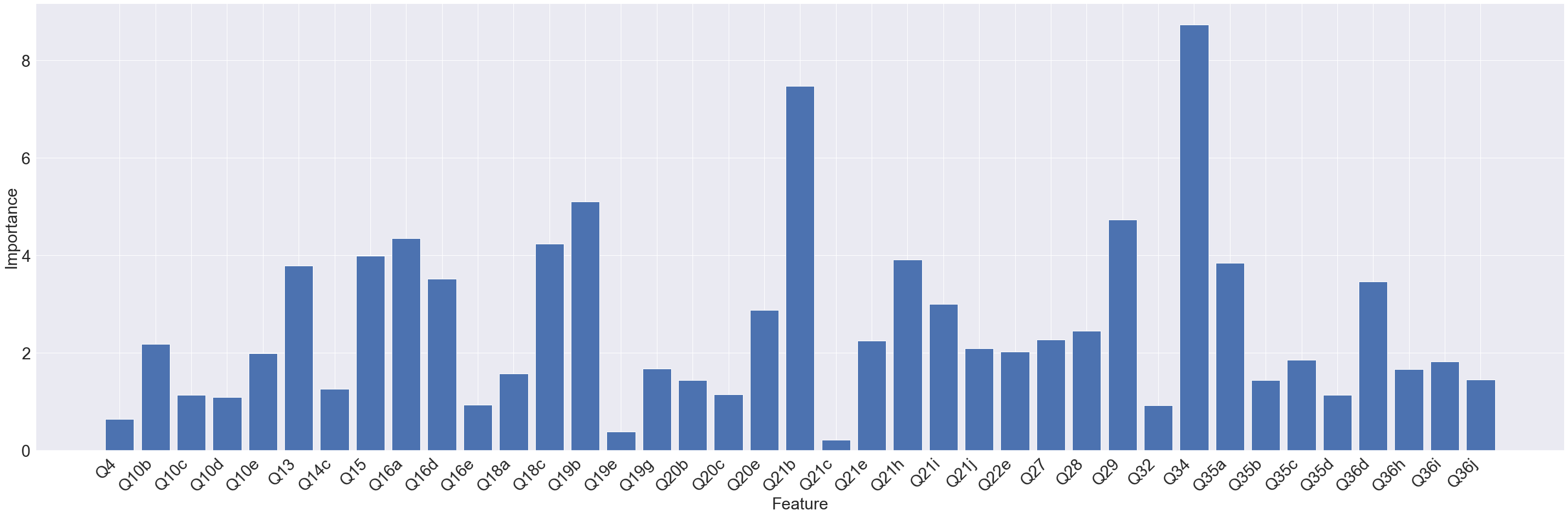}
\label{fig:p6}
\end{figure}

Figure \ref{fig:p5} indicates that the dataset still contains some features that can be removed to reduce the large-p, small-n issues. Based on the feature importance scores in Figure \ref{fig:p5}, we remove features that do not significantly contribute to the model accuracy. We then retrain the CatBoost with 19 trees and obtain the feature importance scores displayed in Figure \ref{fig:p6}. 

\begin{figure}[H]
\caption{The Correlation Heat map for the relationships between some variables in the original dataset (\textit{FD1W})}
\centering
\includegraphics[width=.9\textwidth]{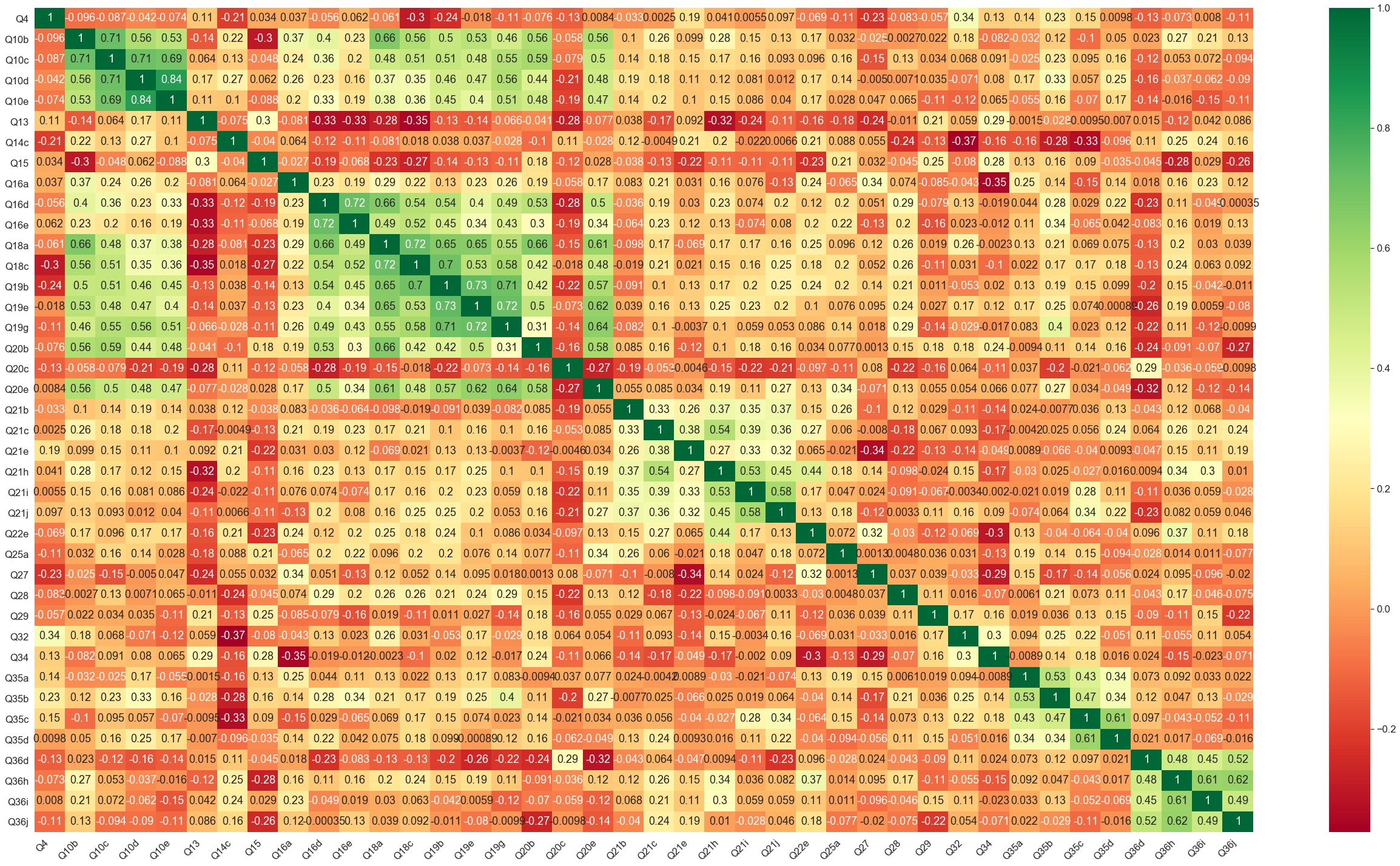}
\label{fig:p7}
\end{figure}

For a better understanding of the relationships among these variables in the original dataset, the \textit{FD1W}, a correlation heat map is provided in Figure \ref{fig:p7}.
Figures \ref{fig:p6} and \ref{fig:p7} indicate that the features most related to Question 25a are Questions 4, 13, 34, 21-b, 29, 19-b, 16-a, 18-a, 35-a, 35-d, 20-e and 36-d, and 36-j (see Figure \ref{fig:p7}). 
\begin{figure}[H]
\caption{The Correlation Heat map for the relationships between the features most related to Question 25a in the original dataset (\textit{FD1W})}
\centering
\includegraphics[width=.7\textwidth]{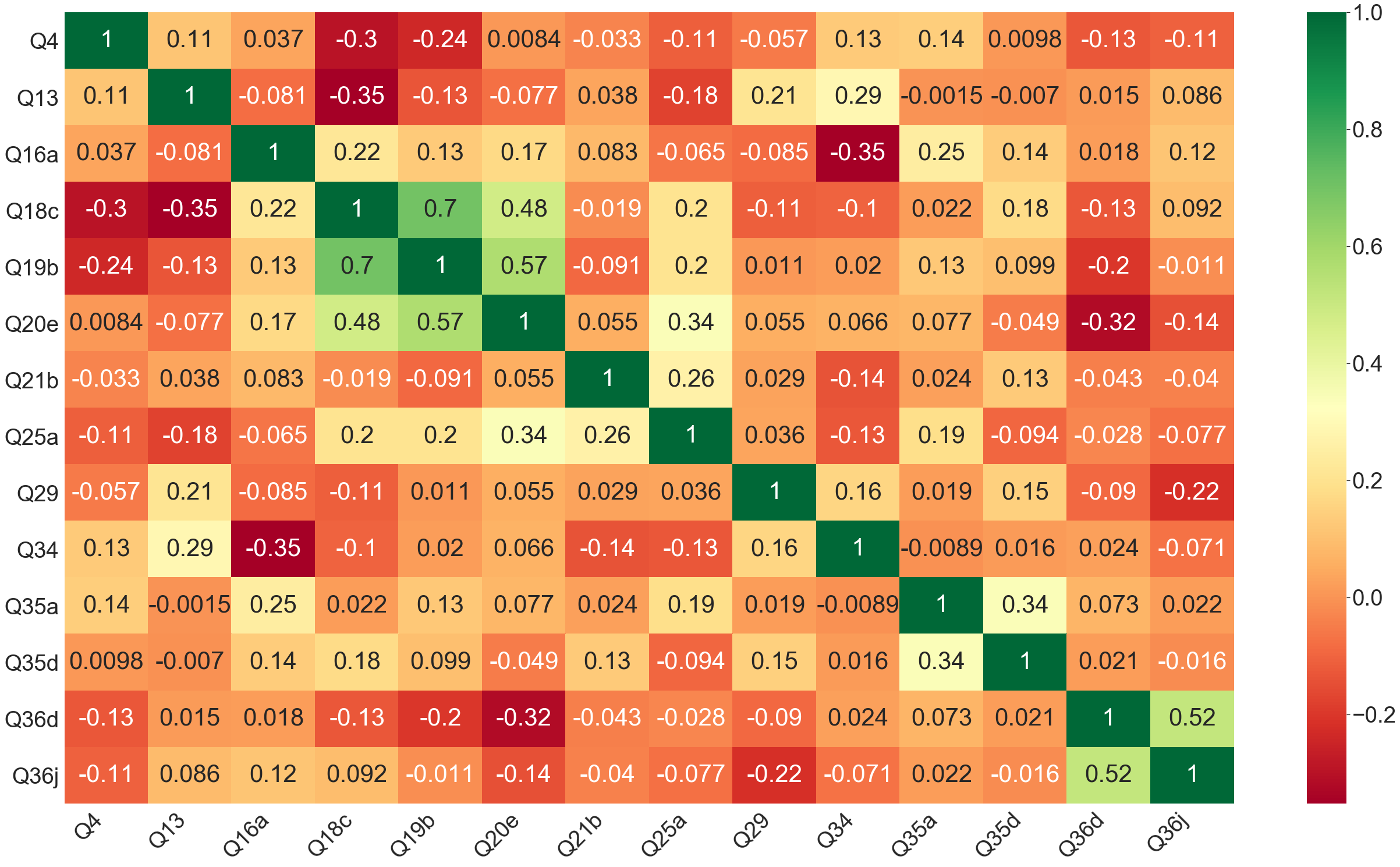}
\label{fig:p8}
\end{figure}
Since the target variable, Question 25, and all variables in Figure \ref{fig:p8} are categorical (nominal or ordinal), we apply a chi-squared test under the null hypothesis $H_0^i:$ Question 25a is independent of Question $i$, where $i \in X=\{$4, 13, 16-a, 18-c, 19-b, 20-e, 21-b, 29, 34, 35-a, 35-d, 36-d, 36-j$\}$ (some important categorical features) on the original dataset \textit{FD1W}. By a significance level of $\alpha= 0.05$, the null hypothesis $H_0^i$ is rejected for all $i \in X$ with a p-value less than 0.01; however, none of supervised multiclass classifiers could obtain a satisfactory accuracy on the \textit{Cleaned-FD1W}, but we determined the features most related to our target variables: Question 25a and Question 25d.

We now retain all important features displayed in Figure \ref{fig:p7} as well as Question 25a through Question 25f, and we then remove the remaining features from the original dataset (\textit{FD1W}). We call the resulting subdataset ``\textit{Q25a-FD1W}.'' 

\subsubsection*{The deletion of rows with missing values from \textit{Q25a-FD1W} and computational results}\label{Section 3.2.2}
First, we remove all rows containing at least one missing value from \textit{Q25a-FD1W}, resulting in a clean 184 by 51 dataset, which includes responses from Afghanistan, Argentina, Bangladesh, Bosnia and Herzegovina, Brazil, Bulgaria, Chile, China, Croatia, Ecuador, Egypt, Georgia, Greece, Guatemala, Hungary, India, Indonesia, Iran, Italy, Japan, Kenya, Kyrgyzstan, Malaysia, Mexico, Mozambique, Nepal, New Zealand, North Macedonia, Oman, Pakistan, Poland, Portugal, Romania, Spain, Thailand, Turkey, United Arab Emirates, United Kingdom, United States of America, and Uzbekistan. 

\begin{figure}[H]
\caption{The distribution of examples among classes of Question 25a in the \textit{Q25a-FD1W} dataset}
\centering
\includegraphics[width=.4\textwidth]{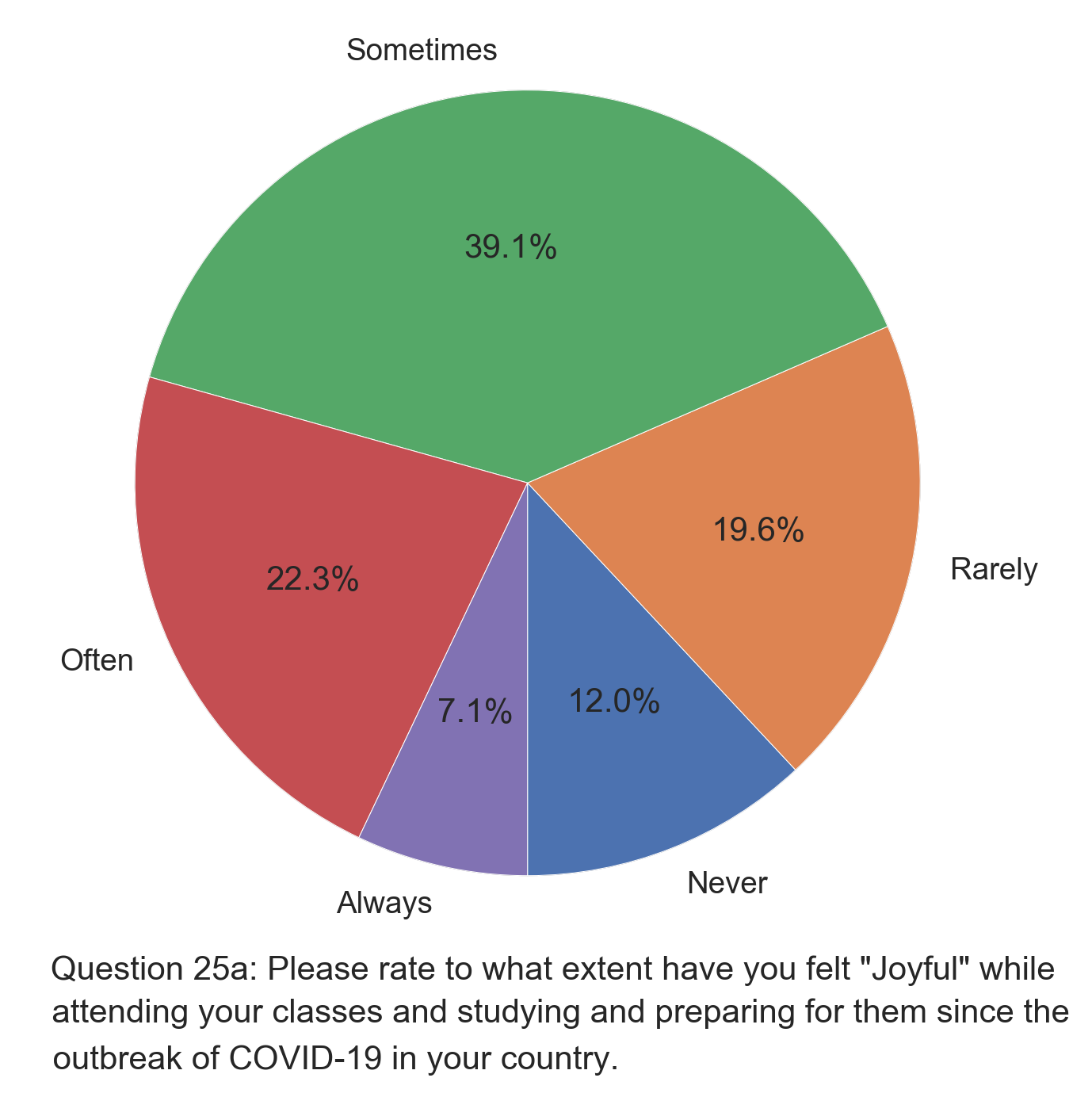}
\label{fig:p9}
\end{figure}

As Figure \ref{fig:p9} illustrates, there is a significant difference in the distribution of the cases in classes of Question 25a in the \textit{Q25a-FD1W} dataset, which means that the dataset is biased toward the class “Sometimes,” which may cause poor performance of supervised machine learning models. One way to resample datapoints to train a more accurate machine learning model is Synthetic Minority Oversampling Technique (SMOTE) \cite{SMOTE}.

\begin{figure}[H]
\caption{Feature scores of Random Forest with 100 trees and depth equals 10 on the \textit{Q25a-FD1W} dataset}
\centering
\includegraphics[width=.9\textwidth]{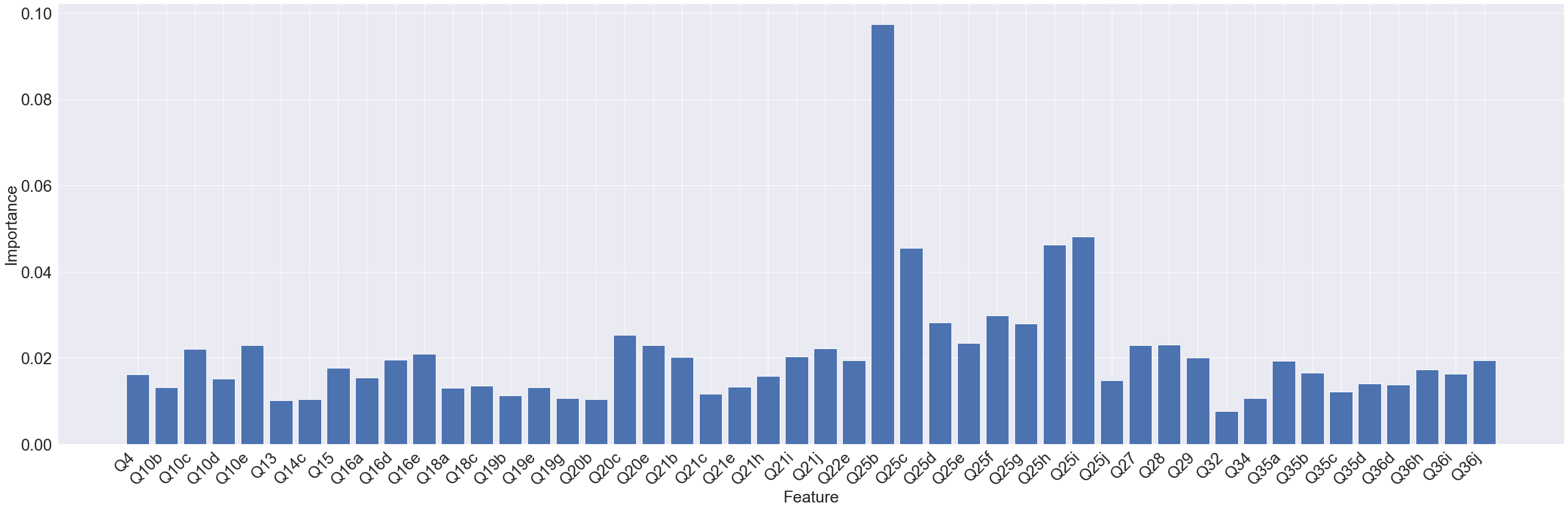}
\label{fig:p11}
\end{figure}

We first split the data into training-test sets (75\%-25\%), and then train a Random Forest containing 100 trees (the depth of each tree is 10). Figure \ref{fig:p11} displays feature importance scores of the model. 

By deleting of rows with missing values from the \textit{FD1W}, we omit important information. In the following subsection, we change our method for dealing with missing values.
\subsubsection*{Dealing with missing values of \textit{Q25a-FD1W} by means of KNN imputation method and computational results}\label{Section 3.2.3}
\begin{figure}[H]
\caption{Feature scores of Random Forest with 200 trees and depth equals 40 on the \textit{Q25a-FD1W} dataset after KNN imputation method and SMOTE are applied}
\centering
\includegraphics[width=.9\textwidth]{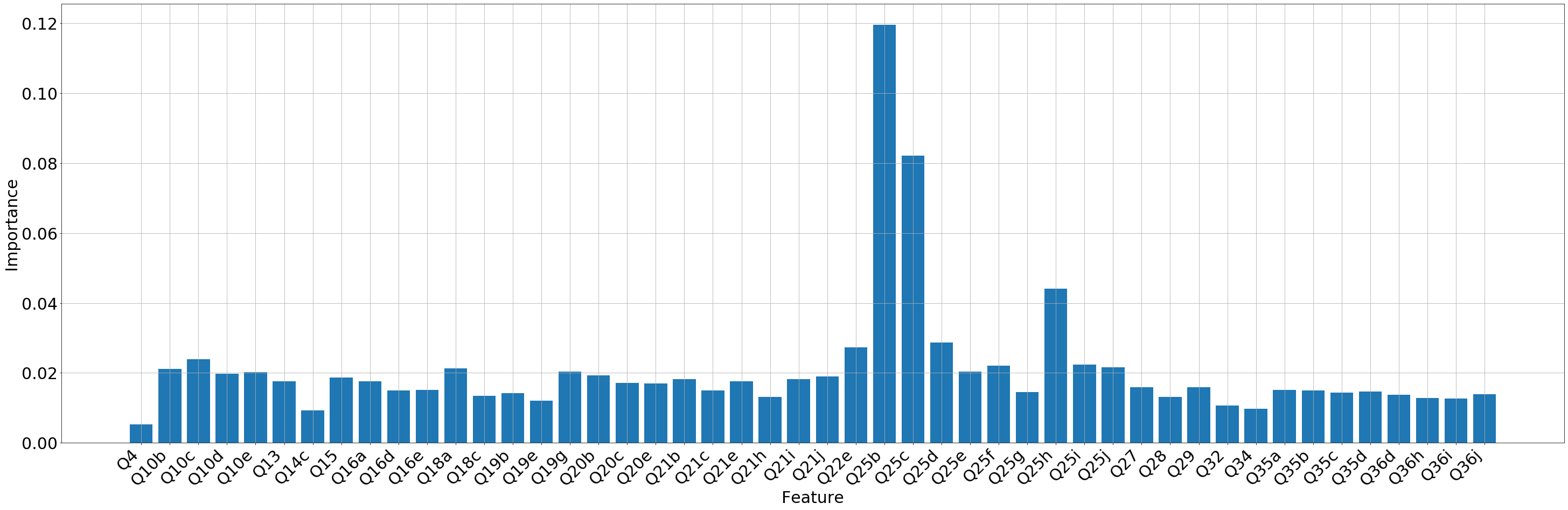}
\label{fig:p26}
\end{figure}

In this subsection, we apply KNN imputation method on the ``\textit{Q25a-FD1W}'' dataset, which contains 237643 missing values. We first fill missing data by means of KNN imputer, split the data into training-test sets (75\%-25\%), apply SMOTE to the training data set, and finally train a Random Forest with with 200 trees of depth 40. But the model obtains an accuracy less than 60\% on test set. Figure \ref{fig:p26} displays feature importance scores of the model. 
Due to cultural, financial, geographical, and political differences, as well as different approaches to education and different ideas regarding the best philosophy for education in different countries, the utilized supervised machine learning models might not obtain a high accuracy score. To eliminate the aforementioned problem arising from differences among different countries, the next two subsections focus on responses from students in the United States of America.    

\subsubsection*{Machine learning analysis of \textit{US-FD1W-25(a)}}\label{Section 3.2.4}
\begin{figure}[H]
\caption{Distribution of examples among Question 25a classes on the \textit{US-FD1W-25(a)} dataset after KNN imputation method is applied}
\centering
\includegraphics[width=.5\textwidth]{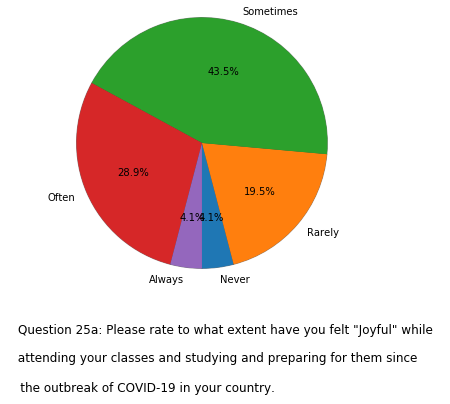}
\label{fig:f1}
\end{figure}

The original dataset \textit{FD1W} contains 392 responses from the US, a 392 by 161 tabular dataset, with 26329 missing values. If we remove all rows with no response to Question 25a, we obtain a 246 by 161 tabular dataset with 6169 missing values; we use the results from previous Subsections, and keep the features most related to Question 25a (and Question 25d), which results in the \textit{US-FD1W-25(a)} dataset. To keep important information and treat missing values appropriately, we use KNN-imputer, which results in a clean (with no missing values) 246 by 49 tabular dataset; however, the distribution of examples among classes of Question 25a is not even (see Figure \ref{fig:f1}).

\begin{figure}[H]
\caption{Feature importance scores of the Random Forest with 200 trees of depth 40 on the \textit{US-FD1W-25(a)} dataset after KNN imputation method and SMOTE are applied}
\centering
\includegraphics[width=.9\textwidth]{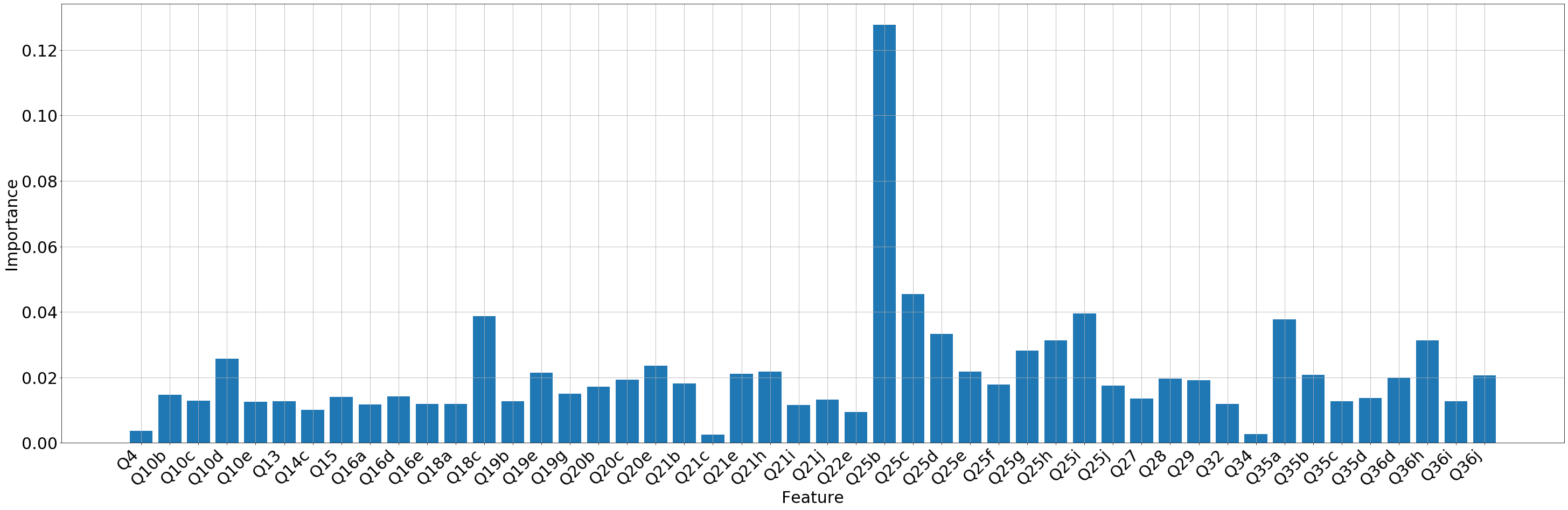}
\label{fig:f3}
\end{figure}

\begin{figure}[H]
\caption{The correlation heat map for some important variables in the \textit{US-FD1W-25(a)} dataset after KNN imputation method}
\centering
\includegraphics[width=\textwidth]{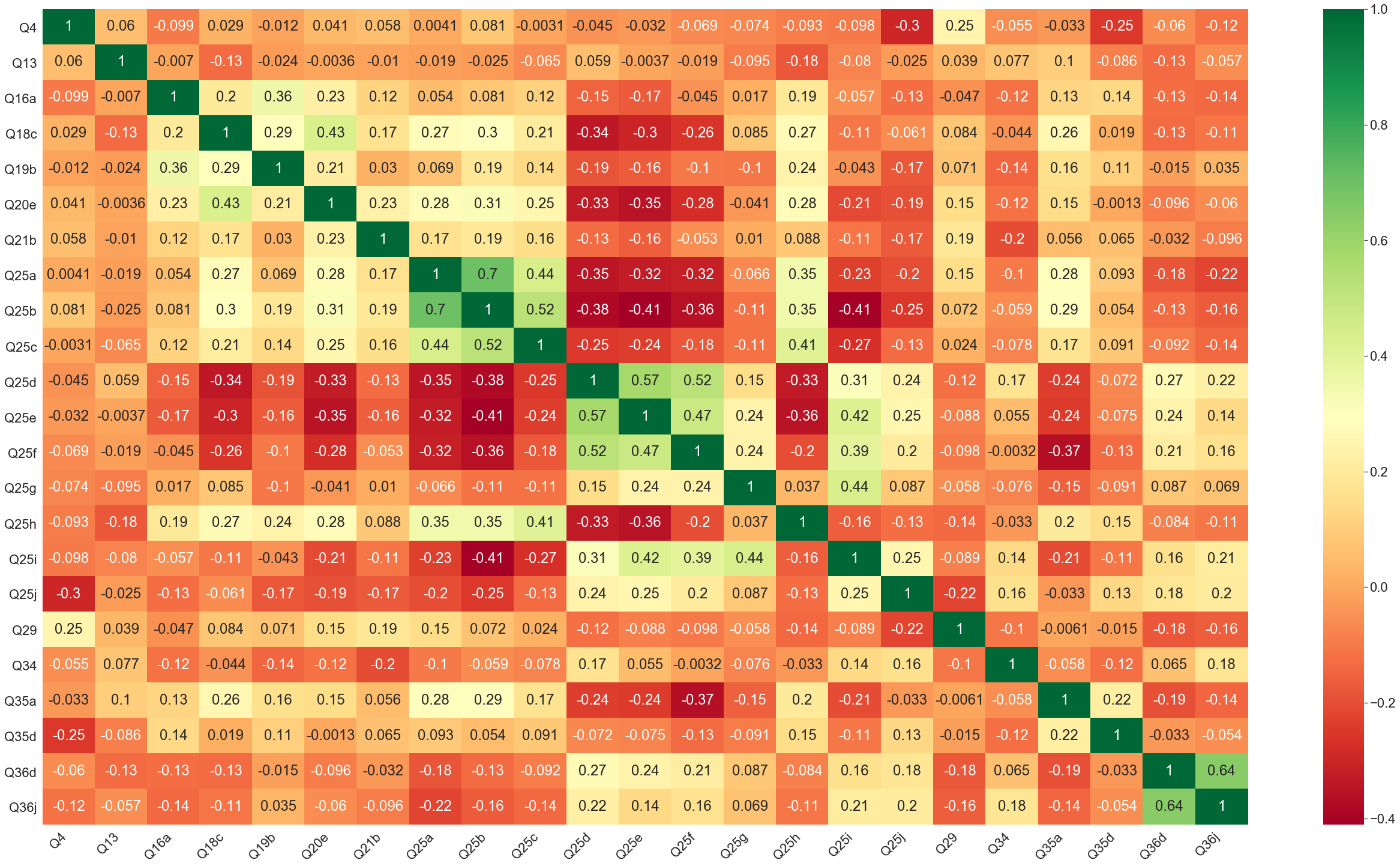}
\label{fig:f4}
\end{figure}

To train robust models, we split the data into training and test sets (75\%-25\%), apply SMOTE on the training data, and then train a Random Forest with 200 trees of depth 40. The Random Forest accuracy score turns out to be still below 65\% with feature importance scores displayed in Figure \ref{fig:f3}.
For a better understanding of relationships among some important features, a correlation heat map for some important variables in the \textit{US-FD1W-25(a)} dataset after KNN imputation method is displayed in Figure \ref{fig:f4}.
Question 25d seems to be a better target variable than Question 25a because it correlates more strongly with the other variables. In the next subsection, we switch from examining Question 25a as the target variable to Question 25d because they appear to be opposites. Hence, we consider Question 25d as the target variable and remove all rows with no response for Question 25d from the US-FD1W, and keep all columns corresponding to Questions 4, 10b, 10c, 10d, 10e, 13, 14c, 15, 16a, 16d, 16e, 18a, 18c, 19b, 19e, 19g, 20b, 20c, 20e, 21b, 21c, 21e, 21h, 21i, 21j, 22e, 25a, 25b, 25c, 25d, 25e, 25f, 25g, 25h, 25i, 25j, 27, 28, 29, 32, 34, 35a, 35b, 35c, 35d, 36d, 36h, 36i, and 36j and remove the remaining columns, then we have 245 rows, 49 columns and 2405 missing values. This subdataset is called \textit{US-FD1W-25(d)}.

\clearpage
% Include only the SI item label in the paragraph heading. Use the \nameref{label} command to cite SI items in the text.
\section*{Data Description.}
\label{Data Description}
To have a better understanding of questions (variables), we categorize them into ten categories as follows:
\begin{itemize}
\item Demographic Information: Questions 1-8
\item Academic Life: Questions 9-18 \& 20
\item Satisfaction with University, Government, and other organizations: Questions 19, 35, 36
\item Access to required materials/skills for online learning: Questions 21 \&22
\item Social Life/Support Network: Questions 23 \& 24
\item Emotions: Question 25
\item Worries/concerns: Question 26
\item Finances: Questions 27 – 34
\item Forced move due to COVID: Question 37
\item Habits/Choices before and since the pandemic: Question 28
\end{itemize}

Below a list of questions, and their descriptions is give. 

% \usepackage[normalem]{ulem}
{
\begin{table}
\begin{tabular}{||p{0.1\textwidth}|p{0.5\textwidth}|p{0.3\textwidth}||}
\hline
Name    & Label & Values                                                                                                                                                                                                                                                                                                                                                                                                                                                                                                                                                                                                     \\
\hline

Q1      & In   which country do you study (in this semester)?                                                                                                                                                                                                                                            &                                                                                                                                                                                                                                                                                                                                       \\
\hline
Q3      & Are you   a citizen of the country in which you study?                                                                                                                                                                                                                                         & 1:   "Yes", 2: "No"                                                                                                                                                                                                                                                                                                                   \\
\hline 
\end{tabular}
\end{table}
\begin{table}
\begin{tabular}{||p{0.1\textwidth}|p{0.5\textwidth}|p{0.3\textwidth}||}
\hline
Name    & Label & Values                                                                                                                                                                                                                                                                                                                                                                                                                                                                                                                                                                                                     \\
\hline
Q11     & Which of these forms of online lectures has been the   most dominant?                                                                                                                                                                                                                          & 1:   "Online in real-time (videoconference)", 2: "Online with a   video recording (not in real-time)", 3: "Online with an audio   recording (not in real-time)", 4: "Online by sending presentations   to students", 5: "Written communication (forums, chat, etc.)"                                                                  \\
\hline
Q12     & {Since on-site classes were cancelled, the   organization of tutorials/seminars and practical classes has changed. Below,   several different forms of online tutorials/seminars and practical classes   are listed. Please assess your level of satisfaction with each form. If you   have no experience with particular form, please select “Not applicable”.}                                                                                                                                                                                                                                                    & \\
\hline
\end{tabular}
\end{table}
\begin{table}
\begin{tabular}{||p{0.1\textwidth}|p{0.5\textwidth}|p{0.3\textwidth}||}
\hline
Name    & Label & Values                                                                                                                                                                                                                                                                                                                                                                                                                                                                                                                                                                                                     \\
\hline
Q14     & Since on-site classes were cancelled, my   supervisions/mentorship has been organized (please select ALL the answers   that are true for you).                                                                                                                                                 &                                                                                                                                                                                                                                                                                                                                       \\
\hline
Q14a    & Via video-call                                                                                                                                                                                                                                                                                 & 1:   "Yes", 0: "No"                                                                                                                                                                                                                                                                                                                   \\
\hline
\end{tabular}
\end{table}
\begin{table}
\begin{tabular}{||p{0.1\textwidth}|p{0.5\textwidth}|p{0.3\textwidth}||}
\hline
Name    & Label & Values                                                                                                                                                                                                                                                                                                                                                                                                                                                                                                                                                                                                     \\
\hline
Q18     & Since on-site classes were cancelled, how satisfied   have you been with the organization of:                                                                                                                                                                                                  &                                                                                                                                                                                                                                                                                                                                       \\
\hline
Q18a    & lectures.                                                                                                                                                                                                                                                                                      & 1:   "Very dissatisfied", 2: "Dissatisfied", 3:   "Natural", 4: "Satisfied", 5: "Very satisfied"                                                                                                                                                                                                                                      \\
\hline
\end{tabular}
\end{table}
\begin{table}
\begin{tabular}{||p{0.1\textwidth}|p{0.5\textwidth}|p{0.3\textwidth}||}
\hline
Name    & Label & Values                                                                                                                                                                                                                                                                                                                                                                                                                                                                                                                                                                                                     \\
\hline
Q19i    & student   counseling services.                                                                                                                                                                                                                                                                 & 1:   "Very dissatisfied", 2: "Dissatisfied", 3:   "Natural", 4: "Satisfied", 5: "Very satisfied"                                                                                                                                                                                                                                      \\
\hline
Q20     & In view of the new teaching and learning environment,   to what extent do you agree with the following statements.                                                                                                                                                                             &                                                                                                                                                                                                                                                                                                                                       \\
\hline
\end{tabular}
\end{table}
\begin{table}
\begin{tabular}{||p{0.1\textwidth}|p{0.5\textwidth}|p{0.3\textwidth}||}
\hline
Name    & Label & Values                                                                                                                                                                                                                                                                                                                                                                                                                                                                                                                                                                                                     \\
\hline
Q21e    & A   printer                                                                                                                                                                                                                                                                                    & 1:   "Never", 2: "Rarely", 3: "Sometimes", 4:   "Often", 5: "Always"                                                                                                                                                                                                                                                                  \\
\hline
Q21f    & Headphones   and microphone                                                                                                                                                                                                                                                                    & 1:   "Never", 2: "Rarely", 3: "Sometimes", 4: "Often",   5: "Always"                                                                                                                                                                                                                                                                  \\
\hline
\end{tabular}
\end{table}
\begin{table}
\begin{tabular}{||p{0.1\textwidth}|p{0.5\textwidth}|p{0.3\textwidth}||}
\hline
Name    & Label & Values                                                                                                                                                                                                                                                                                                                                                                                                                                                                                                                                                                                                     \\
\hline
Q23     & How often have you communicated   with the following people online since the COVID-19 pandemic online?                                                                                                                                                                                         &                                                                                                                                                                                                                                                                                                                                       \\
\hline
Q23a    & Close family member                                                                                                                                                                                                                                                                            & 1:   "Not at all", 2: "Two or three times a month", 3:   "Once a week", 4: "Several times a week", 5: "Once a   day", 6: "Several times a day"                                                                                                                                                                                        \\
\hline
\end{tabular}
\end{table}
\begin{table}
\begin{tabular}{||p{0.1\textwidth}|p{0.5\textwidth}|p{0.3\textwidth}||}
\hline
Name    & Label & Values                                                                                                                                                                                                                                                                                                                                                                                                                                                                                                                                                                                                     \\
\hline
Q23i    & Voluntary organizations                                                                                                                                                                                                                                                                        & 1:   "Not at all", 2: "Two or three times a month", 3:   "Once a week", 4: "Several times a week", 5: "Once a   day", 6: "Several times a day"                                                                                                                                                                                        \\
\hline
Q23j    & Social networks                                                                                                                                                                                                                                                                                & 1:   "Not at all", 2: "Two or three times a month", 3:   "Once a week", 4: "Several times a week", 5: "Once a   day", 6: "Several times a day"                                                                                                                                                                                        \\
\hline
\end{tabular}
\end{table}
\begin{table}
\begin{tabular}{||p{0.1\textwidth}|p{0.5\textwidth}|p{0.3\textwidth}||}
\hline
Name    & Label & Values                                                                                                                                                                                                                                                                                                                                                                                                                                                                                                                                                                                                     \\
\hline
Q24d    & Talk about problems related to   future education.                                                                                                                                                                                                                                             & 1:   "Close family member", 2: "More distant family member",   3: "Close friend", 4: "Someone I live with (e.g.   roommate)", 5: "Neighbours", 6: "Colleague from my   course", 7: "Lecturer", 8:    "Administrative staff", 9: "Voluntary   organizations", 10: "Social networks", 12: "No one"                                      \\
\hline
Q24e    & Talk about problems related to   personal finances.                                                                                                                                                                                                                                            & 1: "Close   family member", 2: "More distant family member", 3:   "Close friend", 4: "Someone I live with (e.g. roommate)",   5: "Neighbours", 6: "Colleague from my course", 7:   "Lecturer", 8:    "Administrative staff", 9: "Voluntary   organizations", 10: "Social networks", 12: "No one"                                      \\
\hline
\end{tabular}
\end{table}
\begin{table}
\begin{tabular}{||p{0.1\textwidth}|p{0.5\textwidth}|p{0.3\textwidth}||}
\hline
Name    & Label & Values                                                                                                                                                                                                                                                                                                                                                                                                                                                                                                                                                                                                     \\
\hline
Q24h    & Be there for you if you would like   to talk about the COVID-19 crisis.                                                                                                                                                                                                                        & 1:   "Close family member", 2: "More distant family member",   3: "Close friend", 4: "Someone I live with (e.g.   roommate)", 5: "Neighbours", 6: "Colleague from my   course", 7: "Lecturer", 8:    "Administrative staff", 9: "Voluntary organizations",   10: "Social networks", 12: "No one"                                      \\
\hline

Q25     & Please rate to what extent have   you felt the following emotions while attending your classes and studying and   preparing for them since the outbreak of COVID-19 in your country.                                                                                                           &                                                                                                                                                                                                                                                                                                                                       \\
\hline
\end{tabular}
\end{table}
\begin{table}
\begin{tabular}{||p{0.1\textwidth}|p{0.5\textwidth}|p{0.3\textwidth}||}
\hline
Name    & Label & Values                                                                                                                                                                                                                                                                                                                                                                                                                                                                                                                                                                                                     \\
\hline
Q26     & How often you have worries about the following   personal circumstances?                                                                                                                                                                                                                       &                                                                                                                                                                                                                                                                                                                                       \\
\hline
Q26a    & Personal physical health                                                                                                                                                                                                                                                                       & 1:   "A little of the time", 2: "Some of the time", 3: "A   good part of the time", 4: "Most of the time", 5: "All of   the time"                                                                                                                                                                                                     \\
\hline
\end{tabular}
\end{table}
\begin{table}
\begin{tabular}{||p{0.1\textwidth}|p{0.5\textwidth}|p{0.3\textwidth}||}
\hline
Name    & Label & Values                                                                                                                                                                                                                                                                                                                                                                                                                                                                                                                                                                                                     \\
\hline
Q27     & When   you take into account your total monthly disposable income (before COVID-19   pandemic) would you say that you can pay the overall costs of your study   (including living costs such as accommodation etc.)?                                                                           & 1:   "Very easily", 2: "Easily", 3: "Quite easily",   4: "With small difficulty", 5: "With difficulty"                                                                                                                                                                                                                                \\
\hline
Q28     & Have   you had a paid job during the current academic year or were you planning on   having a paid job during the current academic year?                                                                                                                                                       & 1:   "Yes, I worked regularly this academic year.", 2: "Yes, I   worked occasionally this academic year.", 3: "No, I have not worked   this academic year but was planning to.", 5: "No, I have not worked   this academic year and was not planning to work."                                                                        \\
\hline
\end{tabular}
\end{table}
\begin{table}
\begin{tabular}{||p{0.1\textwidth}|p{0.5\textwidth}|p{0.3\textwidth}||}
\hline
Name    & Label & Values                                                                                                                                                                                                                                                                                                                                                                                                                                                                                                                                                                                                     \\
\hline
Q35     & Please   assess the satisfaction in dealing with the COVID-19 pandemic of the   following institutions.                                                                                                                                                                                        &                                                                                                                                                                                                                                                                                                                                       \\
\hline
Q35a    & Government                                                                                                                                                                                                                                                                                     & 1:   "Very dissatisfied", 2: "Dissatisfied", 3:   "Natural", 4: "Satisfied", 5: "Very satisfied"                                                                                                                                                                                                                                      \\
\hline
\end{tabular}
\end{table}
\begin{table}
\begin{tabular}{||p{0.1\textwidth}|p{0.5\textwidth}|p{0.3\textwidth}||}
\hline
Name    & Label & Values                                                                                                                                                                                                                                                                                                                                                                                                                                                                                                                                                                                                     \\
\hline
Q36g    & Delayed   tax filing deadline and due date for taxes owed.                                                                                                                                                                                                                                     & 1:   "Not important", 2: "Slightly important", 3: "Moderately   important", 4: "Important", 5: "Very importnant"                                                                                                                                                                                                                      \\
\hline

Q36h    & Deferred   student loan payments.                                                                                                                                                                                                                                                              & 1:   "Not important", 2: "Slightly important", 3:   "Moderately important", 4: "Important", 5: "Very   importnant"                                                                                                                                                                                                                    \\
\hline
\end{tabular}
\end{table}
\begin{table}
\begin{tabular}{||p{0.1\textwidth}|p{0.5\textwidth}|p{0.3\textwidth}||}
\hline
Name    & Label & Values                                                                                                                                                                                                                                                                                                                                                                                                                                                                                                                                                                                                     \\
\hline
Q38f\_1 & Stocked up on   essentials at pharmacy and grocery store                                                                                                                                                                                                                                       & 1:   "Never", 2: "Rarely", 3: "Sometimes", 4:   "Often", 5: "Always"                                                                                                                                                                                                                                                                  \\
\hline
Q38g\_1 & Made a plan for   communicating with family friends and neighbours                                                                                                                                                                                                                             & 1:   "Never", 2: "Rarely", 3: "Sometimes", 4:   "Often", 5: "Always"                                                                                                                                                                                                                                                                  \\
\hline
\end{tabular}
\end{table}
\begin{table}
\begin{tabular}{||p{0.1\textwidth}|p{0.5\textwidth}|p{0.3\textwidth}||}
\hline
Name    & Label & Values                                                                                                                                                                                                                                                                                                                                                                                                                                                                                                                                                                                                     \\
\hline
Q38\_2  & Please assess the   frequency of your habits before and during the COVID-19 pandemic. - During   the COVID-19 pandemic                                                                                                                                                                         &                                                                                                                                                                                                                                                                                                                                       \\
\hline
Q38a\_2 & Washing your   hands                                                                                                                                                                                                                                                                           & 1:   "Never", 2: "Rarely", 3: "Sometimes", 4:   "Often", 5: "Always"                                                                                                                                                                                                                                                                  \\
\hline
\end{tabular}
\end{table}
}

\end{document}